\documentclass{aa}

\usepackage{graphicx,amssymb,amsmath}
\usepackage{natbib}
\usepackage{units}
\usepackage{lipsum}
\usepackage{dcolumn}
\usepackage{wasysym}
\usepackage{hyperref}
\usepackage{txfonts}
\usepackage[T1]{fontenc}
\usepackage[usenames]{color}

\newcommand{\dd}[1]{{\mathrm{d}#1}}
\newcommand{\ee}[1]{{\mathrm{e}^{#1}}}

\newcommand{\sub}[2]{#1_\mathrm{#2}}

\newcommand{\vv}{\mathrm{v}}
\newcommand{\ml}{M}
\newcommand{\Avbar}{\sub{\bar{A}}{V}}
\newcommand{\Avdes}{\sub{\bar{A}}{V,des}}
\newcommand{\Avdis}{\sub{\bar{A}}{V,dis}}

\newcommand{\logten}{\sub{\log}{10}}

\newcolumntype{d}{D{.}{.}{-1}}
\newcolumntype{e}{D{,}{\pm}{-1}}

\newcommand{\gas}{\mathrm{H_2O}}
\newcommand{\ice}{\mathrm{\textrm{s-}H_2O}}
\newcommand{\oxy}{\mathrm{O}}

\newcommand{\orthoground}{\mathrm{H_2O}~1_{10}-1_{01}}
\newcommand{\paraground}{\mathrm{H_2O}~1_{11}-0_{00}}
\newcommand{\paraone}{\mathrm{H_2O}~2_{02}-1_{11}}
\newcommand{\paratwo}{\mathrm{H_2O}~2_{11}-2_{02}}

\newcommand{\EvA}{\tiny{\Circle}}
\newcommand{\EvB}{\tiny{\LEFTcircle}}
\newcommand{\EvC}{\tiny{\CIRCLE}}
\newcommand{\EvD}{\tiny{\CIRCLE\!\LEFTcircle}}
\newcommand{\EvE}{\tiny{\CIRCLE\!\CIRCLE}}
\newcommand{\nEv}{\EvE}

\defcitealias{wal13}{W13}
\defcitealias{alb13}{A13}
\defcitealias{vis11}{V11}

\begin{document}

\authorrunning{Schmalzl et.\ al (2014)}
\titlerunning{The Link Between Water Gas and Ice in Protostellar Envelopes}

\title{Water in Low-Mass Star-Forming Regions with Herschel: The Link Between Water Gas and Ice in Protostellar Envelopes
\thanks{{\it Herschel} is an ESA space observatory with science instruments provided by European-led Principal Investigator consortia and with important participation from NASA.}}

\author{
    M.\ Schmalzl\inst{1}
    \and
    R.\ Visser\inst{2}
    \and
    C.\ Walsh\inst{1}
    \and
    T.\ Albertsson\inst{3}
    \and
    E.F.\ van~Dishoeck\inst{1,5}
    \and
    L.E.\ Kristensen\inst{1,4}
    \and
    J.C.\ Mottram\inst{1}
}

\institute{
    Leiden Observatory, Leiden University, P.O. Box 9513, 2300 RA Leiden, the Netherlands\\
    \email{schmalzl@strw.leidenuniv.nl}
    \and
    Department of Astronomy, University of Michigan, 1085 S.\ University Ave, Ann Arbor, MI 48109-1107, USA
    \and
    Max-Planck-Institut f\"ur Astronomie, K\"onigstuhl 17, 69117 Heidelberg, Germany
    \and
    Harvard-Smithsonian Center for Astrophysics, 60~Garden Street, MS~42, Cambridge, MA 02138, USA
    \and
    Max-Planck-Institut f\"ur Extraterrestrische Physik, Giessenbachstrasse, Garching, Germany
}

\date{Accepted 15/09/2014}

\abstract
{}
{Our aim is to determine the critical parameters in water chemistry and 
the contribution of water to the oxygen budget by observing and modelling
water gas and ice for a sample of eleven low-mass protostars, for
which both forms of water have been observed.}
{A simplified chemistry network, which is benchmarked against
more sophisticated chemical networks, is developed that includes the necessary
ingredients to determine the water vapour and ice abundance profiles in the
cold, outer envelope in which the temperature increases towards the protostar.
Comparing the results from this chemical network to
observations of
water emission lines and previously published water ice column densities,
allows us to probe the influence of various agents (e.g., FUV
field, initial abundances, timescales, and kinematics).}
{The observed water ice abundances with respect to hydrogen
nuclei in our sample are
$\unit[30-80]{ppm}$, and therefore contain only
10--30\% of the volatile oxygen budget of $\unit[320]{ppm}$. The keys to 
reproduce this result are a low initial water ice abundance after the pre-collapse phase
together with the fact that atomic oxygen cannot freeze-out and form water ice in
regions with $\sub{T}{dust}\gtrsim\unit[15]{K}$. This requires
short prestellar core lifetimes $\lesssim\unit[0.1]{Myr}$. The
water vapour profile is shaped through the interplay of
FUV photodesorption, photodissociation, and freeze-out. The water vapour
line profiles are an invaluable tracer for the FUV photon flux
and envelope kinematics.}
{The finding that only a fraction of the oxygen budget is locked in
water ice can be explained either by a short pre-collapse time of
$\lesssim\unit[0.1]{Myr}$ at densities of $\sub{n}{H}\sim\unit[10^4]{cm^{-3}}$,
or by some other process that resets the initial water
ice abundance for the post-collapse phase. A key for the understanding of the
water ice abundance is the binding energy of atomic oxygen on ice.}


\maketitle

\section{Introduction}

In cold clouds ($\sim\unit[10]{K}$), water is predominantly found
on dust grains in the form of
water ice. Its main formation pathway is in-situ: atomic oxygen
is accreted from the gas phase onto the grain
surface and is successively hydrogenated to form water ice
\citep[e.g.][]{tie82, hir98, miy08, iop08, iop10, mok09, oba09, cup10, dul10}.
Water vapour can form in the gas
phase through ion-molecule chemistry at low temperatures and through
neutral-neutral reactions at high temperatures
\citep[for recent review, see][]{dis13}.
The gas and ice phases are linked through freeze-out from the gas
phase onto dust grains, and through thermal and
non-thermal desorption of ice back into the gas phase. Because
water ice formation is efficient and starts already in molecular clouds
prior to collapse, most models of pre- and protostellar evolution
turn the bulk of the available oxygen into water ice in the cold parts
of the cores \citep[e.g.,][]{aik08,hol09,cas12}.

Water ice can be observed through infrared absorption of vibrational
bands superposed on the continuum of an embedded young stellar object
or a background star. Surveys of large samples of low- and high-mass
protostars as well as background sources reveal
typical water ice column density ratios
$\sub{N}{\ice}/\sub{N}{H}\sim5\times10^{-5}$
\citep[e.g.,][]{smi93,gib04,pon04,boo04,boo08,boo11,whi01,whi07,whi13,obe11}. Here
$N_{\rm H}$ is the column density of hydrogen nuclei, inferred either
from the silicate optical depth or the colour excess toward the star.
Although these values can vary by up to a factor of 2, the implication
is that water ice contains only a small fraction, $<$20\%, of the
overall interstellar oxygen abundance with respect to
hydrogen atoms of $5.75\times 10^{-4}$
\citep{prz08}. Water gas can lock up a large fraction of oxygen,
but only in hot gas where high-temperature neutral-neutral reactions
drive most of the oxygen not contained in refractory grains into water
\citep[e.g.,][and references therein]{dis11}. Indeed, in cold regions the water gas
abundance has been found to be very low, $10^{-10}-10^{-8}$, as
inferred from observations with the \textit{Submillimeter Wave
Astronomy Satellite} (SWAS) and subsequent missions
\citep{sne00,ber00,cas12}.

While the overall picture of high ice and low gas-phase water
abundances in cold clouds appears well established, there are only a few
sources for which both ice and gas have been observed along the same
line of sight. Using the \textit{Infrared Space Observatory}, the column
densities of ice and warm water vapour have been determined through
mid-infrared absorption lines toward a dozen high-mass infrared-bright
sources \citep{dis96,boo03a}. The warm water absorption lines
originate in the inner envelope where the dust temperature is above
the water ice sublimation temperature, which is of the order of
$\sim\unit[100]{K}$ \citep[e.g.,][]{fra01,bur10},
and column densities comparable to those of water ice have been
found. The infrared absorption data of warm water have subsequently
been combined with SWAS submillimetre emission lines of cold water of
the same high-mass sources to infer the gas-phase water abundance
profile in both the cold and the warm gas \citep{boo03b}. Using
different trial abundance structures applied to physical models of the
sources constrained from continuum data, a jump in the gas-phase water
abundance of $\sim$4 orders of magnitude from cold to warm regions was
established. Using a standard ice abundance of $10^{-4}$, the
ice column in these models was found to be a factor
of 3--6 above the observed values. Although no full gas-grain model was
adopted, even this simple empirical analysis showed difficulties in
getting the water gas and ice chemistry to be consistent.

With the increased sensitivity of ground- and space-based infrared and
submillimetre instrumentation, the combined study of water gas and ice
can now be extended to low-mass protostars. In particular, ESO-VLT
3$\mu$m and {\it Spitzer} mid-infrared \textit{water ice}
spectra exist for about 50
infrared-bright low-mass protostars \citep[e.g.,][]{bro05,boo08},
and the \textit{Herschel Space Observatory} \citep{pil10}
has observed submillimetre lines of \textit{water vapour} of a
comparable sized sample. Here we investigate the overlapping set of
eleven low-mass protostars.

The goal of this paper is to constrain the relative importance of the main
processes that shape the water vapour and ice abundance profiles in
the cold parts of protostellar envelopes. To this end, a
\textit{Simplified Water Network} (SWaN) is developed, which is a
dedicated tool that only incorporates the key processes that control
the water gas and ice abundances in regions with temperatures
$\lesssim\unit[100]{K}$. This model is then combined with
physical envelope models from
\citet{kri12}. Together these data provide insight into the cold water
chemistry as well as the puzzle as to why water ice occupies only a
minor fraction of the available oxygen \citep{whi10}.

The paper is structured as follows. In Sect.~\ref{s-obs} we
introduce the \textit{Herschel} HIFI observations of our sample, and
give a brief overview of the observations and existing physical models
of the sources. In Sect.~\ref{s-abundances} we introduce our
simplified network (with the benchmarking results against full
networks shown in the Appendix in Sect.~\ref{s-benchmark}) and analyse its
sensitivity to key parameters. In Sect.
\ref{s-comparison} we compare the models with the observations of
water ice column densities and \textit{Herschel} spectra, which is
followed by a discussion on the important parameters in
Sect.~\ref{s-discussion} and the conclusions in
Sect.~\ref{s-conclusion}.

\section{Observations and Physical Models}
\label{s-obs}

\begin{table*}
\caption{\label{t-01} List of protostellar cores and their envelope properties. }
\centering
\begin{tabular}{l|c|cddd|c|c|e|cccc}
\hline \hline Name & $\sub{T}{bol}$\tablefootmark{a} & \multicolumn{1}{c}{$\sub{n}{H,0}$\tablefootmark{a}} & \multicolumn{1}{c}{$r_0$\tablefootmark{a}} & \multicolumn{1}{c}{$\sub{r}{env}$\tablefootmark{a}} & \multicolumn{1}{c|}{$\alpha$\tablefootmark{a}} & \multicolumn{1}{c|}{$\sub{N}{H}$\tablefootmark{a}} & \multicolumn{1}{c|}{$\sub{N}{H}^*$\tablefootmark{a}} & \multicolumn{1}{c|}{$\sub{N}{\ice}$} & \multicolumn{4}{c}{Water Transition} \\
 & (K) & \multicolumn{1}{c}{(cm$^{-3}$)} & \multicolumn{1}{c}{(au)} & \multicolumn{1}{c}{($10^3$ au)} & & \multicolumn{1}{c|}{(cm$^{-2}$)} & \multicolumn{1}{c|}{(cm$^{-2}$)} & \multicolumn{1}{c|}{\tiny{($\unit[10^{18}]{cm^{-2}}$)}} & o0 & p0 & p1 & p2 \\ \hline
L\,1527    & 44 & 1.8(8) & 5.4 & 4.6 & 0.9 & 1.4(23) & 1.2(23) & 4.7,1.3\tablefootmark{d} & $\bullet$ & $\bullet$ & $\bullet$ & $\bullet$ \\
IRAS\,15398   & 52 & 3.9(9) & 6.2 & 2.7 & 1.4 & 8.2(23) & 6.2(23) & 14.8,4.0\tablefootmark{b} & $\bullet$ & $\bullet$ & $\bullet$ & $\bullet$ \\ \hline
L1551-IRS5   & 94 & 1.4(9) & 28.9 & 14.1 & 1.8 & 7.4(23) & 4.4(23) & 10.9,0.2\tablefootmark{c} & $\bullet$ & $\bullet$ & $\bullet$ & $\bullet$ \\
TMC1     & 101 & 1.7(8) & 3.7 & 5.0 & 1.1 & 4.8(22) & 3.9(22) & 7.9,0.2\tablefootmark{c} & $\bullet$ & $\bullet$ & $\bullet$ &   \\
HH\,46    & 104 & 7.1(8) & 28.5 & 16.8 & 1.6 & 4.9(23) & 3.3(23) & 7.8,0.8\tablefootmark{b} & $\bullet$ & $\bullet$ & $\bullet$ & $\bullet$ \\
TMC1A    & 118 & 1.0(9) & 7.7 & 6.7 & 1.6 & 2.0(23) & 1.3(23) & 5.3,0.2\tablefootmark{c} & $\bullet$ & $\bullet$ & $\bullet$ &   \\
RCrA-IRS5   & 126 & 2.2(7) & 10.0 & 10.0 & 0.8 & 4.9(22) & 4.5(22) & 3.6,0.3\tablefootmark{b} & $\bullet$ &   &   &   \\
TMR1     & 133 & 8.2(8) & 8.8 & 6.8 & 1.6 & 1.8(23) & 1.2(23) & 7.4,0.3\tablefootmark{c} & $\bullet$ & $\bullet$ & $\bullet$ & $\bullet$ \\
L\,1489    & 200 & 4.1(8) & 8.4 & 6.2 & 1.5 & 9.8(22) & 6.9(22) & 4.3,0.5\tablefootmark{b} & $\bullet$ & $\bullet$ & $\bullet$ & $\bullet$ \\
HH\,100-IRS   & 256 & 3.5(6) & 15.5 & 15.5 & 0.5 & 4.9(22) & 4.7(22) & 2.5,0.2\tablefootmark{b} & $\bullet$ &   &   &   \\
RNO\,91    & 340 & 2.7(8) & 6.6 & 6.0 & 1.2 & 9.9(22) & 8.1(22) & 4.2,0.4\tablefootmark{b} & $\bullet$ & $\bullet$ & $\bullet$ & $\bullet$ \\
\hline
\end{tabular}
\tablefoot{The evolutionary stage is determined by the bolometric temperature,
$\sub{T}{bol}$. The horizontal line separates Class~0 and Class~I sources,
using $\sub{T}{bol}=\unit[70]{K}$ as criterion for separation \citep{che95}.
The envelope is characterised by a power-law density distribution with
volume density at the inner edge, $\sub{n}{H,0}$, inner radius, $\sub{r}{0}$,
envelope radius $\sub{r}{env}$, and power-law exponent $\alpha$
(Equation~\ref{e-01}). The overall column density along the line-of-sight
from the envelope edge towards the core centre is denoted by $\sub{N}{H}$,
whereas $\sub{N}{H}^*$ represents the column density in the water-freezeout
zone, i.e., regions where water is mostly in icy form (see Sect.~\ref{s-repmodel}
for more details). $\sub{N}{\ice}$ is the observed water ice column density
\citep{boo08,zas09,aik12}. The abbreviations for the water transition lines
are: (o0) ortho ground state $\orthoground$; (p0) para ground state $\paraground$;
(p1) $\paraone$; (p2) $\paratwo$. The bullet symbol ($\bullet$)
indicates that this transition has been observed.\\
\tablefoottext{a}{Envelope parameters from \citet{kri12}};
Water ice column densities are from
\tablefoottext{b}{\citet{boo08},}
\tablefoottext{c}{\citet{zas09}, and}
\tablefoottext{d}{\citet{aik12}.}
}
\end{table*}

\subsection{Observations}

Our observations of water vapour lines were obtained within the
framework of the \textit{Herschel} key project
\textit{Water in Star-forming Regions with Herschel} \citep[WISH;][]{dis11}.
The sub-sample analysed here was specifically selected to
contain observations of both water vapour and water ice column
densities. Thus, our target list is limited to eleven infrared-bright
low-mass protostars (two Class~0, nine Class~I; Table~\ref{t-01}).
The observed column densities of water ice
range from $\unit[3.0\times10^{18}$ to $1.5\times10^{19}]{cm^{-2}}$
\citep{boo08,zas09,aik12}. Since the water ice column density is estimated by
observing water ice absorption in MIR spectra towards the embedded protostar, these
column densities are always upper limits for the ice content of the envelope.
Other contributions can be provided by, e.g., foreground clouds or disks.

{\it Herschel} observations are performed
with the \textit{Heterodyne Instrument for the Far-Infrared}
\citep[HIFI][]{gra10}, which provides line profiles of cold water
vapour down to an unprecedented sensitivity and spatial resolution and
opens up the spectral window to observe emission lines from higher
excited levels.
For all eleven protostars, observations have been performed in the ortho-ground
state transition $\orthoground$ at $\unit[557]{GHz}$.
For most sources there are also data of the para ground-state
$\paraground$ at $\unit[1113]{GHz}$, plus two
additional excited transitions -- $\paraone$ at
$\unit[988]{GHz}$, and $\paratwo$ at
$\unit[752]{GHz}$. A list of observing dates and IDs is found in
Table~\ref{t-09} in the Appendix. It should be noted that there are two
more sources with observations of both the water ice and water vapour,
but they are excluded from our sample due to contamination with foreground
absorption (Elias\,29) and outflow emission (GSS30-IRS), which do not allow
to draw any conclusions from their \textit{Herschel} observations.

\begin{figure*}[tb]
    \includegraphics{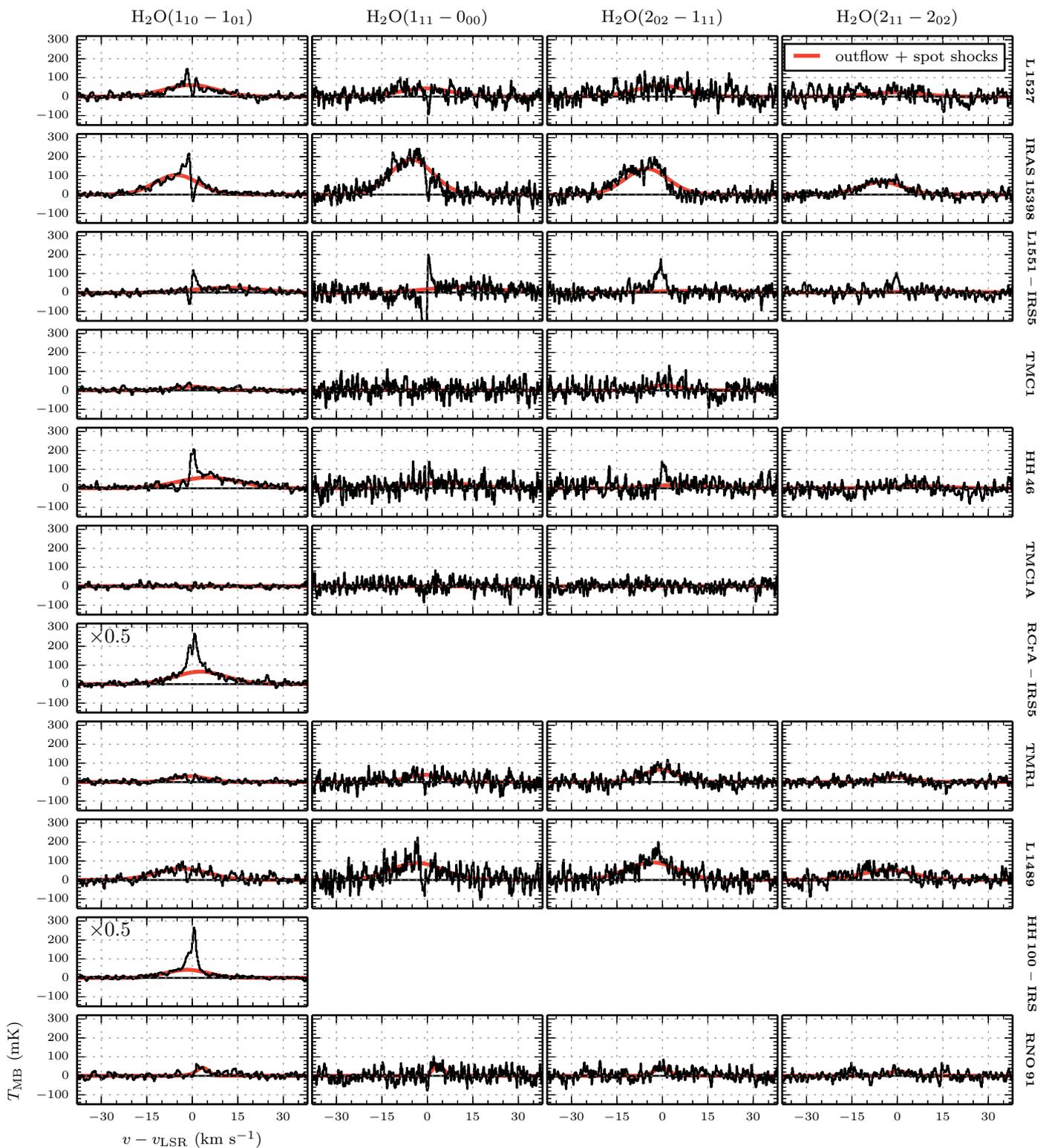}
    \caption{\label{f-01} Overview of the continuum-subtracted
    \textit{Herschel} observations of the water transitions for all
    protostellar cores in our sample. The fit of the outflow and spot
    shock-emission, which is subtracted in the data presented in Fig.~\ref{f-05}
    and Fig.~\ref{f-06}, is shown as a smooth curve.}
\end{figure*}

For the data reduction, the two polarisations (H, V) are combined and
corrected for main beam efficiency as described in
\citet{kri12}. For low-mass sources, the HIFI
lines are dominated by broad features due to the outflow, which is
not of interest here. A Gaussian decomposition following
\citet{mot14} has been be used to subtract the contribution of
the water emission from the outflows and spot shocks.
The continuum-subtracted spectra alongside
the best-fits for the non-envelope contribution are depicted in
Fig.~\ref{f-01}.

\subsection{Physical models}
\label{s-physmod}

To analyse the water gas and ice data, a physical model of the
protostellar envelope is needed which specifies the density and
temperature profiles of the protostellar envelope. Spherically
symmetric model fits for each source have been made by \citet{kri12}.
Following the procedure by \citet{jor02} the density profile
is characterised by a power-law density
structure of the form
\begin{align}
\label{e-01}
\sub{n}{H}(r)=\sub{n}{H,0}\,\left(\frac{r}{r_0}\right)^{-\alpha}.
\end{align}
The
dust temperature has been determined self-consistently
by performing a
full continuum radiative transfer calculation assuming a central
source with the observed luminosity as input. The best fitting values
of $\alpha$, envelope mass, and envelope extent have been obtained by comparison
with the spectral energy distribution and sub-millimetre continuum images of the
sources. The envelope is truncated at the point where either the dust temperature
is $<\unit[10]{K}$ or the hydrogen number density is
$<\unit[2\times10^4]{cm^{-3}}$, whichever comes first.
These points mark the transition between
the envelope and the ambient cloud, which in our simulations is assumed to be
chemically inert and devoid of water.
The gas temperature is mostly coupled to the dust temperature, but
following the approach by \citet{bru12} we take into account that in regions
with elevated UV radiation field $\sub{T}{gas}>\sub{T}{dust}$.

Fractional abundances with respect to hydrogen nuclei of
various species $X_i\equiv n_i/\sub{n}{H}$ are
specified at each radius, $r$, measured from the centre of the
core. Alternatively, the visual extinction, $A_{\rm V}$, measured from
the outer edge of the envelope, $\sub{r}{env}$, can be used to describe the chemistry,
acknowledging the fact that substantial changes in abundance profiles
are introduced with the attenuation of the interstellar radiation field (ISRF).
The extinction in the radial direction is obtained from
the models through $A_{\rm V}(r)=\int_{r}^{\sub{r}{env}} \sub{n}{H}(r^\prime)
\dd{r^\prime} /\unit[1.9\times 10^{21}]{cm^{-2} mag^{-1}}$ where the
conversion factor in the denominator is taken from the empirical
determination by \citet{boh78} and \citet{rie85}.

\section{Model Abundances in Protostellar Envelopes}
\label{s-abundances}

\subsection{Simplified Water Network (SWaN)}
\label{s-swan}

\begin{figure}[tb]
    \includegraphics{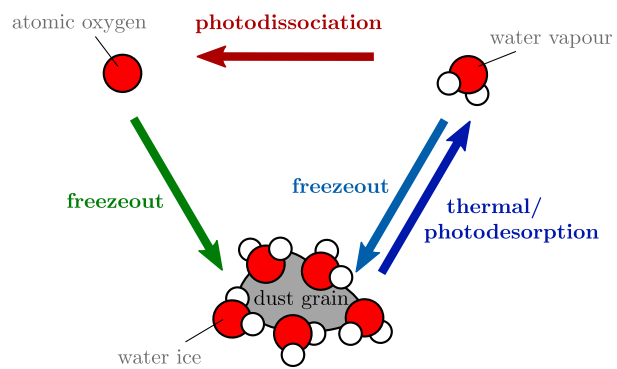}
    \caption{\label{f-02} Simplified chemical network with three
    main components: \textit{i)}~atomic oxygen (O), \textit{ii)}~water ice on the
    grain surface ($\ice$), and \textit{iii)}~water vapour ($\gas$).
    See Sect.~\ref{s-swan} for a more detailed description.}
\end{figure}

With the aid of observations of both water vapour and ice and a
prescription for the temperature and density structure of the sources,
we attempt to understand the key processes that shape the water
abundance profiles in the cold parts of the protostellar envelopes.
For this purpose a \textbf{S}implified \textbf{Wa}ter \textbf{N}etwork
(SWaN) is developed, which is reduced to a minimum set of ingredients
and reaction channels needed to reliably determine the abundance
structure of water vapour and water ice.
A comparison of SWaN to more sophisticated chemical networks (\citealt{vis11},
\citetalias{vis11}; \citealt{alb13}, \citetalias{alb13};
\citealt{wal13}, \citetalias{wal13}) can be
found in Appendix~\ref{s-benchmark}.

Previous studies used step- or drop-abundance profiles
for water vapour \citep{her12,cou12,cou13}. These profiles are characterised
by distinct regions of constant abundance. In the case of drop-abundance profiles,
these are the inner region ($T\gtrsim\unit[100]{K}$)
with $\sub{X}{\gas}\sim10^{-5}$, the outer region
with $\sub{X}{\gas}\sim10^{-8}$, and a photodesorption layer at the envelope edge
with $\sub{X}{\gas}\sim10^{-7}$ \citep{cou12}. The location of the transition
between the inner and outer region is placed at $T\sim\unit[100]{K}$, but
the extent of the photodesorption layer is a-priori unknown and has to be
estimated by other means. The advantage of SWaN over
these phenomenological profiles is that we have full control over
the parameters that shape the abundance profile, and the photodesorption layer
and its extent comes as a natural consequence of the interplay of the physical
and chemical processes. The strength of these processes and
their relative importance can be assessed, which allows us to study the link between
water gas and ice. Our approach is similar to the work
by \citet{cas12} and \citet{ket14} on water vapour abundance
profiles in prestellar cores, except our study with SWaN focuses on
the understanding of the connection between water gas and ice in the cold,
outer envelope of protostellar cores, which are characterised by
a temperature increase towards the centre.

Our simplified network, SWaN, consists of three species
(water ice on the grain surfaces, water vapour, and atomic oxygen),
which are connected by four reaction channels
(Fig.~\ref{f-02}). In the following description of the
chemical network, all reaction rates $R_i$ are in units of
$\unit{cm^{-3}\,s^{-1}}$, and correspond to the number of
atoms/molecules that are transformed through a particular reaction channel
per unit volume and unit time.

Water ice can desorb through {FUV photodesorption} at a rate
\begin{align}
 \label{e-02}
 \sub{R}{phdes}=4\,\sub{\sigma}{H}\,\sub{n}{H}\,\sub{F}{FUV}\,\sub{Y}{pd},
\end{align}
where $\sub{\sigma}{H}$ is the grain cross section per hydrogen nuclei, and is given by
$\sub{\sigma}{H}=\sigma_\mathrm{gr}\,\sub{n}{gr}/\sub{n}{H}$ (with
the grain cross section $\sigma_\mathrm{gr}=\pi a_\mathrm{gr}^2$,
the grain radius $\sub{a}{gr}$, and the grain volume density $\sub{n}{gr}$),
$F_\mathrm{FUV}$ is the flux of FUV photons at the
grain surface \footnote{Equation~(\ref{e-02})
considers an isotropic ISRF, i.e.\ photons from all directions reach
the grains, whereas 1D chemical models \citep[e.g.][]{hol09}
consider UV photons from the ISRF coming only from one direction. In
these models, it is then the grain cross section instead of the full
surface that is able to capture UV photons. Consequently, our equation
includes an additional factor of 4
(Appendix~\ref{a-av}).} in units of $\unit{s^{-1}cm^{-2}}$,
and $Y_\mathrm{pd}$ is the photodesorption
yield. Following recent lab results \citep{obe09,ber12}, FUV
photodesorption is treated as a zeroth-order process (i.e., molecules can only
desorb from the top few layers). \citet{obe09} give a photodesorption yield of
\begin{align}
 \label{e-03}
 \sub{Y}{pd} = \sub{Y}{pd,0}\,\sub{\theta}{\ml},
\end{align}
where $\sub{Y}{pd,0}=10^{-3}$ is
the photodesorption yield for thick ice, and
$\sub{\theta}{\ml}$ is the monolayer coverage factor
\begin{align}
 \label{e-04}
 \sub{\theta}{\ml}=1 - \ee{-\ml/l},
\end{align}
where $l = 0.6$ is the diffusion length. $\ml$ is the number of monolayers, which
is given by
\begin{align}
 \label{e-05}
 \ml=\frac{\sub{n}{\ice}/\sub{n}{gr}}{4\sigma_\mathrm{gr}\,N_\mathrm{s}},
\end{align}
where $\sub{N}{s}=\unit[1.5\times10^{15}]{cm^{-2}}$ is the density of sites
on the grain surface \citep{has92}. This approach is supported by molecular dynamics
simulations \citep{and08,ara10}. In reality, FUV
absorption of water ice results not only in H$_2$O but also OH
molecules escaping from the ice \citep{and08,obe09}, but the latter
channel is not taken into account explicitly.

In our model, the flux of FUV photons consists of two components
$\sub{F}{FUV}=\sub{F}{FUV,isrf}+\sub{F}{FUV,cr}$, where the first term
is the contribution of FUV photons from the ISRF, and the second term
reflects the secondary UV field caused by cosmic rays interacting with
molecular hydrogen \citep{pra83}. The flux of ISRF FUV photons
penetrating the ice mantle at a certain depth into the cloud can be
calculated via
\begin{align}
 \label{e-06}
 F_\mathrm{FUV,isrf}=F_0\,\sub{G}{isrf}\,\ee{-1.8\,\Avdes}
\end{align}
with $\sub{G}{isrf}$ as the scaling factor for the standard
ISRF photon flux $F_0$, and $\Avdes$ as the
spherically averaged extinction for photodesorption
(Appendix~\ref{a-av}). The flux of secondary FUV photons induced by
cosmic rays is independent of extinction, and results in an
isotropic, constant FUV photon flux
\begin{align}
 \label{e-07}
 F_\mathrm{FUV,cr}=F_0\,\sub{G}{cr}
\end{align}
with $\sub{G}{cr}$ depending on the assumed energy distribution of the
cosmic rays \citep{she04}. Other
mechanisms such as direct cosmic-ray desorption have no
significant impact \citep{hol09}, and are therefore not included in
SWaN. Chemical desorption using the excess energy produced by water
ice formation has also been proposed as a desorption mechanism
\citep{dul13} but is still poorly understood qualitatively and
quantitatively and is therefore neglected.

Desorption of water ice from the grain surface can also occur through thermal desorption
\citep[e.g.,][]{vis11} at a rate
 \begin{align}
 \label{e-08}
 \sub{R}{thdes}=4\,\sub{\sigma}{H}\,\sub{n}{H}\,\sub{N}{s}\,\sub{\theta}{\ml}\,\sub{\nu}{\gas}\,\ee{-\sub{T}{b,\gas}/\sub{T}{\rm dust}},
 \end{align}
where $\sub{\nu}{\gas}$ is the lattice vibrational frequency of a water molecule in its
binding site, $T_\mathrm{b,\gas}$ is the binding energy of water expressed as
temperature, and $\sub{T}{dust}$ is the dust temperature. For reasons of simplicity,
the same monolayer coverage $\sub{\theta}{\ml}$ as for the
photodesorption is also adopted for thermal desorption.
The lattice vibrational frequency of water
$\sub{\nu}{\gas}=\unit[2.8\times10^{12}]{Hz}$ is calculated through
the harmonic oscillator approach of \citet{has92}.

Water vapour can be photodissociated through FUV photons, again taking into
account contributions from both the ISRF and the CR-induced secondary field,
at a rate
\begin{align}
 \label{e-09}
 \sub{R}{phdis}=\left(\sub{G}{isrf}\,\ee{-1.7\,\Avdis}+\sub{G}{cr}\right)\,\sub{k}{phdis}\,n_\gas,
\end{align}
where $\sub{k}{phdis}$ is the unshielded
photodissociation rate of water in a FUV field with $\sub{G}{isrf}=1$,
and $\Avdis$ is the spherically averaged mean
extinction through the envelope for photodissociation (see
Appendix~\ref{a-av}). It is assumed that all the water, which is
photodissociated, is turned into atomic oxygen, essentially leaving
out the intermediate product OH. The slightly different
exponential dependence on extinction in Equation~(\ref{e-06}) vs.\
(\ref{e-09}) arises in the FUV absorption
spectrum of water ice, which is shifted to higher energies by about $\unit[1]{eV}$
compared with water vapour \citep{and08}.

Water ice is formed directly through the freeze-out of water
vapour, or indirectly through freeze-out of atomic oxygen. In this
second step, intermediate steps through O$_2$ are ignored, but atomic oxygen
is instantaneously hydrogenised to form water ice. The
benchmarking (Appendix~\ref{s-benchmark}) reveals that the formation
timescale of water ice can be longer than the freeze-out timescale
of atomic oxygen, but also that the abundance profiles
of water vapour and water ice are only marginally affected.
For species $i$, freeze-out occurs at a rate of
\begin{align}
 \label{e-10}
 R_{\mathrm{fr},i}=\sub{\sigma}{H}\,\sub{n}{H}\,{\rm v}_i\,n_i\,S_{\!i}
\end{align}
where $n_i$ is the number density, ${\rm v}_i$ is the thermal velocity, and
$S_{\!i}$ the sticking probability. For water vapour a sticking
probability of unity is assumed. For atomic oxygen we introduce an effective
sticking probability
$\sub{S}{\!O}$, which is determined by the balance between
freeze-out and thermal desorption. We take the relative reaction rates
$\sub{k}{fr,O}=\sub{\sigma}{H}\,\sub{n}{H}\,{\rm v}_i$ and
$\sub{k}{thdes}=\sub{\nu}{\gas}\,\ee{-\sub{T}{b,\gas}/\sub{T}{\rm dust}}$ to
determine $\sub{S}{\!O}=1-\sub{k}{thdes}/\sub{k}{fr,O}$. For the
protostellar cores in our sample this generally means that in regions
with $\sub{T}{dust}\gtrsim\unit[15]{K}$ the thermal desorption rate
is higher than the freeze-out rate, i.e., oxygen atoms cannot freeze out,
and the formation of water ice through the atomic oxygen route is
inhibited.

The number densities $\sub{n}{O}(t)$, $\sub{n}{H_2O}(t)$, and
$\sub{n}{\ice}(t)$ are determined by a set of three differential equations:
\begin{align}
 \frac{\dd{\sub{n}{\ice}}}{\dd{t}} &= \left(\sub{R}{fr,\oxy} + \sub{R}{fr,\gas}\right) \,\, - \,\,\left(\sub{R}{phdes} + \sub{R}{thdes}\right),\\[2ex]
 \frac{\dd{\sub{n}{\gas}}}{\dd{t}} &= \left(\sub{R}{phdes} + \sub{R}{thdes}\right) \,\, - \,\,\left(\sub{R}{fr,\gas} + \sub{R}{phdis} \right),\\[2ex]
 \frac{\dd{\sub{n}{\oxy}}}{\dd{t}} &= \sub{R}{phdis}\,\, - \,\,\sub{R}{fr,\oxy}.
\end{align}
These are solved with the aid of the \textsc{Python} function
{\tt odeint}, which is part of the {\tt
 scipy.integrate}\footnote{http://www.scipy.org} package and makes
use of the \textsc{Fortran} library {\tt odepack}. The standard model parameters are
summarized in Table~\ref{t-02}.

All models use the same
parameters as in Table~\ref{t-02}, but nevertheless show
significant differences, both among
each other and with SWaN. In spite of these uncertainties, the derived abundance
structures for water vapour and water ice are robust in all models. In
contrast, atomic oxygen in our simple network is very different from
the detailed networks and only serves as a proxy for other
oxygen-bearing species within the full water-chemistry network
(e.g., OH, H$_2$O$_2$).

\subsection{Water ice and gas in a representative model}
\label{s-repmodel}

\begin{table}[tb]
\caption{\label{t-02}Standard parameters for the chemistry network.}
 \centering{
 \begin{tabular}{lll}
 \hline \hline parameter & value & reference \\ \hline
 $\sub{a}{gr}$ & $\unit[0.1]{\mu m}$ & (1) \\
 $\sub{X}{gr}$ & $6.5\times10^{-13}$ & (1) \\
 $\sigma_\mathrm{H}$ & $\unit[2.0\times10^{-22}]{cm^2}$ & (1) \\
 $\sub{T}{b,H_2O}$ & $\unit[5530]{K}$ & (2) \\
 $\sub{T}{b,O}$ & $\unit[800]{K}$ & (3) \\
 $\sub{Y}{pd,0}$ & $10^{-3}$ & (4) \\
 $\sub{k}{phdis}$ & $\unit[8.0\times10^{-10}]{s^{-1}}$ & (5) \\
 $\sub{F}{0}$ & $\unit[10^8]{cm^{-2}s^{-1}}$ & (6) \\
 $\sub{G}{isrf}$ & 1.0 \\
 $\sub{G}{cr}$ & $10^{-4}$ & (7) \\ \hline
 \end{tabular}
}
\tablebib{(1)~\citet{ber95};
(2)~\citet{bur10};
(3)~\citet{tie87};
(4)~\citet{obe09};
(5)~\citet{dis06};
(6)~\citet{hab68};
(7)~\citet{she04}
}
\end{table}

Using SWaN as described in the previous section,
the abundance profiles of water vapour and water
ice can be determined. The overall volatile oxygen abundance, i.e., the
oxygen not contained in silicate grains, is taken to be
$\sub{X}{O,ISM}=3.2\times 10^{-4}$,
or $\unit[320]{ppm}$\footnote{ppm: parts per million or $10^{-6}$}, as determined for
diffuse clouds \citep{mey98}. This is the amount of oxygen that can
cycle between the various forms of oxygen-containing gas-phase
and ice species. Of particular interest for this work is the cold, outer
part of the envelope where water is frozen out on the dust grains,
henceforth denoted as the \textit{water freeze-out zone}.

\begin{table}
\centering
\caption{\label{t-03} Abundances of water vapour ($\sub{X}{\gas}$),
water ice ($\sub{X}{\ice}$), atomic oxygen ($\sub{X}{\oxy}$), and the sum
of these three species ($\sub{X}{O,SWaN}$) after various pre-collapse times
$\sub{t}{pre}$ as predicted from the dark cloud model with
$\sub{T}{dust}=\unit[10]{K}$, $\sub{A}{V}=\unit[10]{mag}$,
$\sub{n}{H}=\unit[2\times10^4]{cm^{-3}}$ (\citetalias{wal13}).}
\begin{tabular}{l|ddd|d}
 \hline \hline \multicolumn{1}{c|}{$\sub{t}{pre}$} & \multicolumn{1}{c}{$\sub{X}{\gas}$} & \multicolumn{1}{c}{$\sub{X}{\ice}$} & \multicolumn{1}{c|}{$\sub{X}{\oxy}$} & \multicolumn{1}{c}{$\sub{X}{O,SWaN}$} \\
    \multicolumn{1}{c|}{(Myr)} & \multicolumn{1}{c}{(ppm)} & \multicolumn{1}{c}{(ppm)} & \multicolumn{1}{c|}{(ppm)} & \multicolumn{1}{c}{(ppm)} \\ \hline
 0.01 & 0.1 & 1.9 & 312.8 & 314.8\\
 0.1 & 0.3 & 19.4 & 250.0 & 269.7\\
 1.0 & 0.2 & 102.3 & 61.2 & 163.7\\
 10.0 & 0.1 & 226.7 & 2.3 & 229.1 \\ \hline
\end{tabular}
\end{table}

Our model consists of two stages:
\begin{itemize}
  \item The \textit{pre-collapse} or prestellar phase, where a full chemical
network under the assumption of dark cloud conditions (\citetalias{wal13} with
$\sub{T}{dust}=\unit[10]{K}$,
$\sub{A}{V}=\unit[10]{mag}$, $\sub{n}{H}=\unit[2\times10^4]{cm^{-3}}$) sets the
initial molecular abundances for the second step. Table~\ref{t-03} lists
abundances of water vapour, water ice, and atomic oxygen for different
pre-collapse times. The freeze-out of atomic oxygen is closely connected to
the steady rise of the water ice abundance. At times $\unit[0.1-1.0]{Myr}$,
a considerable amount of oxygen is also found in other oxygen bearing
species (mainly CO), reducing the amount of oxygen within the water chemistry
network. At $\sub{t}{pre}>\unit[1]{Myr}$, oxygen returns into the water
network and water ice then
becomes the only considerable oxygen reservoir.
  \item The second stage is the so-called \textit{post-collapse} phase,
where the abundance structure is modelled
with SWaN on a static envelope with the temperature and density structure
from \citet{kri12}.
\end{itemize}

In
our representative model a pre-collapse time of $\sub{t}{pre}=\unit[0.1]{Myr}$
is chosen, which is motivated by our finding that a rather short
pre-collapse time is required to match the observed low water ice abundances
(see Sect.~\ref{s-res_ice}).
The initial abundance of oxygen within SWaN at that timestep is
$\sub{X}{O,SWaN}=\sub{X}{\gas}+\sub{X}{\ice}+\sub{X}{\oxy}=\unit[270]{ppm}$.
The remaining $\sub{X}{O,ISM}-\sub{X}{O,SWaN}=\unit[50]{ppm}$ can be attributed to
oxygen locked up in CO, s-CO and other oxygen species not contained in the
simple network. The post-collapse timescale in this representative model
is chosen to be $\sub{t}{post}=\unit[1.0]{Myr}$, which is about
the time when equilibrium is reached at all radii. As fiducial values for the
FUV fluxes in this simulations, $\sub{G}{isrf}=1$ and $\sub{G}{cr}=10^{-4}$ are
used.

\begin{figure}[tb]
 \includegraphics{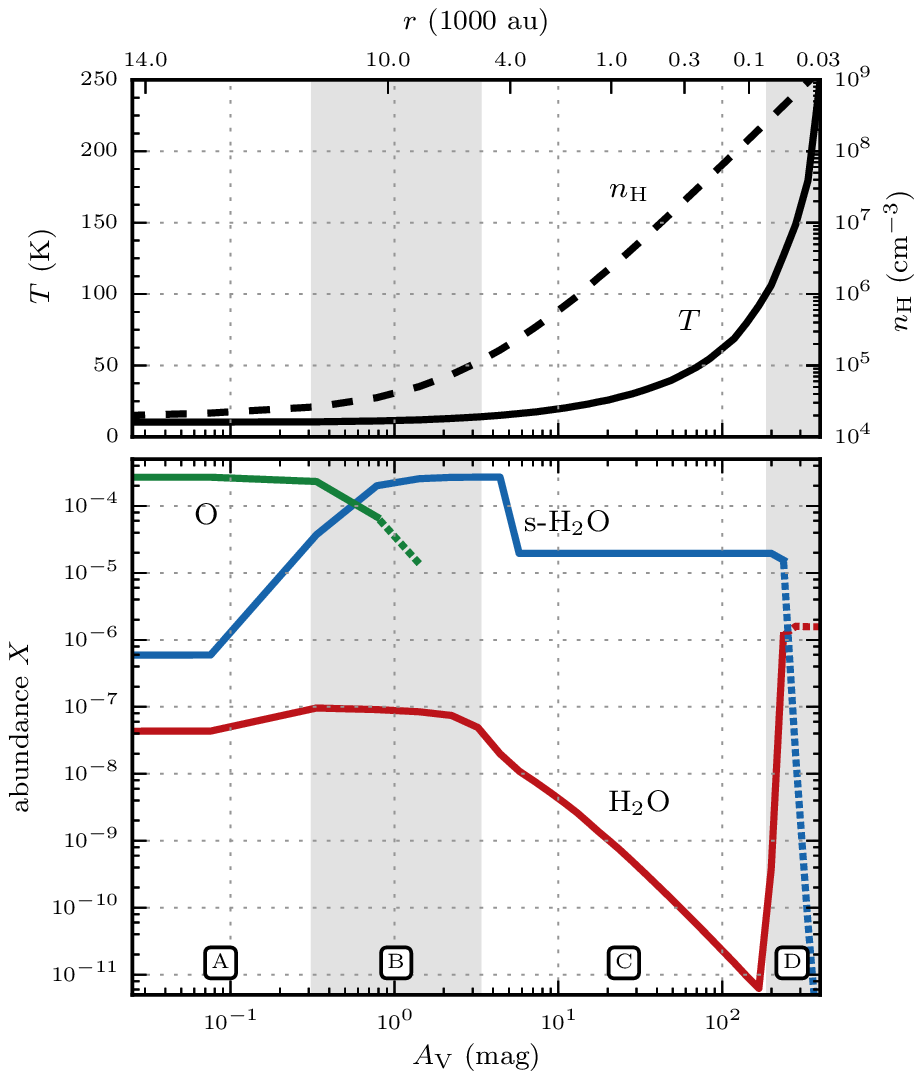}
 \caption{\label{f-03} The \textit{top panel} shows
the hydrogen volume density (dashed) and the gas/dust
temperature (solid) profiles as a function of radial
extinction $\sub{A}{V}$ and radius $r$ for L1551-IRS5 for default
values of the FUV fluxes ($\sub{G}{isrf}=1$, $\sub{G}{cr}=10^{-4}$).
The \textit{bottom panel}
depicts the abundance structure of water ice ($\ice$) and
water vapour ($\gas$) with respect to total hydrogen as
modelled with SWaN after $t=\unit[1]{Myr}$. The benchmarking of SWaN
showed that the abundance of atomic oxygen is not reliably determined
deeper into the cloud, which is why $X(\oxy)$ is not shown
beyond $\sub{A}{V}\gtrsim\unit[2]{mag}$. The different regions
(A-D) are discussed in more detail in the text. It should be noted that
in this and subsequent figures, the edge of the envelope is on the left
and the protostar on the right-hand side.}
\end{figure}

\begin{figure*}[tb]
 \includegraphics{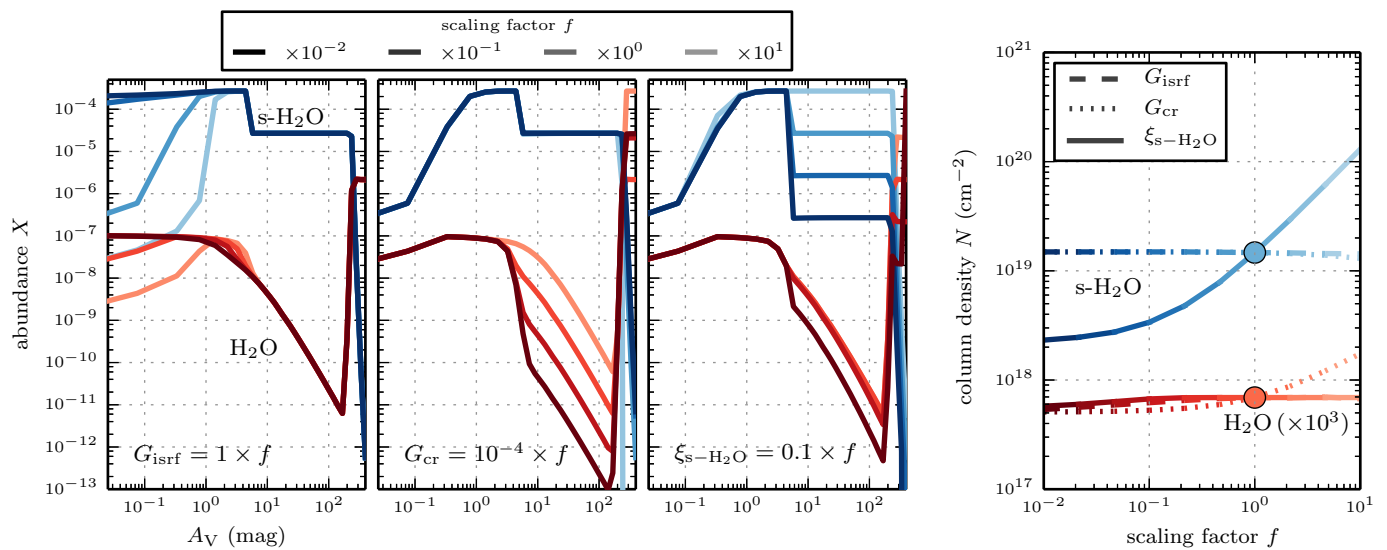}
 \caption{\label{f-04} The dependence of water abundances and
column densities of L1551-IRS5 on the FUV-ISRF field ($\sub{G}{isrf}$),
CR-induced FUV field
($\sub{G}{cr}$), and initial fractional abundance of water ($\sub{\xi}{\ice}$).
The three panels on the left show the abundance profiles of water vapour (red)
and water ice (blue) upon changing one parameter, keeping
the other two parameters at their respective fiducial values
($\sub{G}{isrf}=1$, $\sub{G}{cr}=10^{-4}$, $\sub{\xi}{\ice}=0.1$). The different
shadings represent different scaling factors $f$. The right
panel shows the column density in the water freeze-out zone
as a function of the parameter scaling. The colour shades represent
the respective profile from the left panel. The filled circles represent the
column density with all parameters exhibiting their fiducial level.}
\end{figure*}

The abundance profiles
for water vapour, water ice and atomic oxygen
vs.\ the radial extinction $\sub{A}{V}$ are shown
in Fig.~\ref{f-03} for L1551-IRS5, which is chosen as a representative
core from our sample.The overall structure of the \textit{water vapour} profile
can be roughly separated into four different regions, which
are denoted as Regions~A to D. Depending on the choice
of parameters and the envelope temperature and density structure
the extent of these regions can vary.

The outermost layer (Region A) of the protostellar envelope is dominated by
atomic oxygen. Water ice and water vapour are only
present in trace amounts, since
the impinging ISRF efficiently photodissociates water molecules, but due to
the long freeze-out timescale in the tenuous envelope, the regeneration of
water ice is significantly slower. The attenuation of the ISRF deeper into the core
leads to a build up of water ice and water vapour. When the grains are fully
covered with water molecules (in this model run at around
$\sub{A}{V}\sim\unit[0.7]{mag}$), the water vapour
abundance reaches a plateau (Region B). The extinction value at which this
plateau is reached strongly depends on $\sub{G}{isrf}$
(cf.\ Sect.~\ref{s-paramscan_gasice})
The reaction network in these two outermost regions is characterised by the cycle
$\oxy\rightarrow\ice\rightarrow\gas\rightarrow\oxy$.

That changes deeper into
the cloud (Region C). Due to the increased number densities in concert with
decreased FUV fluxes, the freeze-out of water vapour starts to
outrun the photodissociation. The reaction network reduces to
$\gas\leftrightarrows\ice$, where the water abundances are determined by
the balance between freeze-out of water vapour and photodesorption of water ice.
At $\sub{A}{V}\gtrsim\unit[4-5]{mag}$ (assuming standard values for $\sub{G}{isrf}$
and $\sub{G}{cr}$), the CR-induced FUV field is the main contributor of
FUV photons, which maintain an approximately constant
number density of gas phase water. Since the hydrogen volume
density in this region spans almost three orders of magnitude,
the water vapour abundance
drops considerably from the outer towards the inner edge of Region~C.
Water is predominantly in the form of water ice. For most of our sources,
$50\%$ of the pencil beam water ice column density can be found within the
central $\unit[20-80]{au}$, which is typically around 1\% of the envelope
extent.

The water ice abundance profile shows a
characteristic hump at $\sub{A}{V}\sim\unit[2-3]{mag}$, which is the result of
the freeze-out threshold for atomic oxygen. Only in the oxygen freeze-out zone
($\sub{T}{dust}\lesssim\unit[15]{K}$), can atomic oxygen be transformed
further into water ice. Typically, $\sim5-15\%$ of the column density
in the water freeze-out zone is also part of the oxygen freeze-out zone.
However,
deep in the envelope ($\sub{T}{dust}\gtrsim\unit[15]{K}$)
the water ice abundance remains constant at its initial value.
No water ice can form, but there are also no mechanisms
which can considerably decrease its abundance.

Even deeper in the envelope,
the dust temperature reaches the sublimation temperature of water ice,
which consequently marks the boundary
of the water freeze-out zone. With SWaN we do not attempt to model any
abundances beyond this point
(Region D). Thermal desorption only plays a role in Region D. Nevertheless,
this reaction channel is included in our simplified network, since it allows us
to determine the extent of the water freeze-out-zone.
In our simulations, the limit of the water freeze-out zone is
defined as the radius at which the water ice abundance drops
by three orders of magnitude from its plateau abundance. This is typically
at temperatures of $\sub{T}{dust}=\unit[110-140]{K}$, and its location can
-- depending on the envelope density and temperature structure -- vary by up to a
few au compared to the $\unit[100]{K}$-radius, which is generally the
first order assumption for the sublimation radius.

The water vapour abundance profiles have mostly reached
an equilibrium stage at $\sub{t}{post}=\unit[0.01]{Myr}$.
In the outer envelope (Region A)
this is established through the high photodesorption and photodissociation rates,
and deeper into the envelope (Region C) through the high freeze-out rates. In the
transition region between Regions B and C the situation is a bit different.
The hydrogen number densities are $\sim\unit[10^5]{cm^{-3}}$
(which results in low freeze-out rate),
but the ISRF FUV photon flux is already considerably attenuated (which results in
low photo-rates). Therefore, in this region
it takes much longer ($\sim\unit[0.1-1.0]{Myr}$) to reach an equilibrium stage.
However, the discussed timescales can change upon the choice of initial conditions.
It should be noted that this region does not affect the total ice or gas column density
(Sect.~\ref{s-dis_ice}), but it does affect the water emission line profiles
(Sect.~\ref{s-paramscan_gas}).

\subsection{Dependence on Model Parameters}

\subsubsection{Water Abundances and Column Densities}
\label{s-paramscan_gasice}

\begin{figure*}[tb]
 \includegraphics{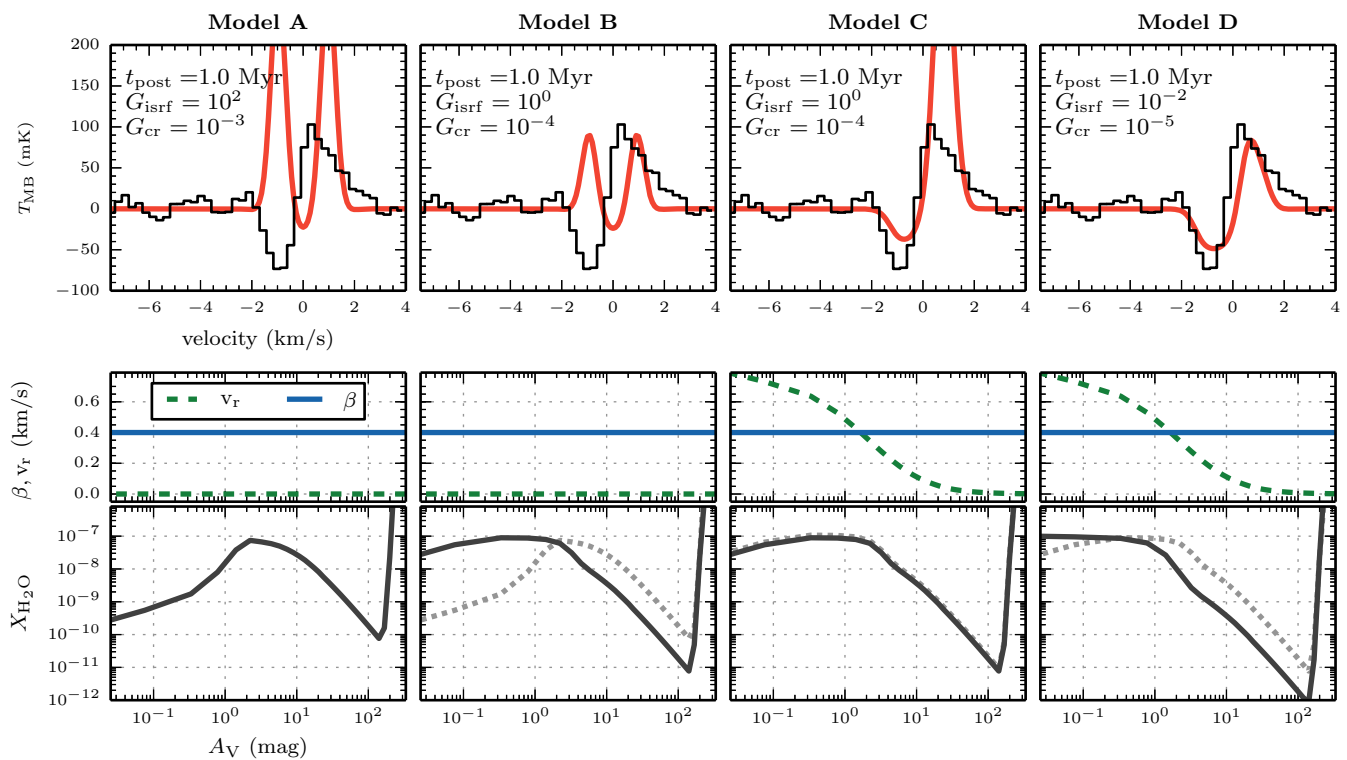}
 \caption{\label{f-05} Synthetic $\mathrm{H_2O}(1_{10}-1_{01})$
observations for L1551-IRS5, assuming different parameters for the
FUV fluxes ($\sub{G}{isrf}$ and $\sub{G}{cr}$), and
velocity profiles (radial velocity $\sub{\vv}{r}$ and the turbulent
broadening Doppler-$\beta$)
in a sequence from Model A to D. The
top row shows the synthetic spectra (red)
alongside the observations (black), which have
been shifted by the systemic velocity. The middle
row depicts the assumed velocity profile, and the bottom row shows the
water abundance profile. For comparison, the abundance profile of the
preceding model is shown as dashed line.}
\end{figure*}

The focus of this work is to determine the key factors that shape the
water gas and ice abundance profiles and regulate their column
densities, with particular attention to its dependence on the FUV
photon fluxes and initial abundances. In these simulations, the initial
abundances of a pre-collapse time of $\unit[0.1]{Myr}$ are used
(Table~\ref{t-03}). The initial overall
oxygen abundance in the system is $\sub{X}{O,SWaN}=\unit[270]{ppm}$. About 7\%
of the oxygen is found in the form of water ice, but for demonstrative
purposes we use a fiducial value for the initial fraction of water ice of
$\sub{\xi}{\ice}=10\%$ to allow easier scaling in order of magnitudes.
The fiducial values for the FUV fields
are $\sub{G}{isrf}=1$ and $\sub{G}{cr}=10^{-4}$.
In three different runs, each parameter is scaled by
a factor of $f=10^{-2}-10^1$, leaving the other two parameters
at their fiducial values.

An enhanced $\sub{G}{isrf}$ can be
due to UV from a nearby bright star but potentially also from
from fast shocks related to the protostellar jet impinging on the envelope.
An enhanced ISRF can lead to an increased gas temperature \citep[e.g.,][]{hol97}.
A lower than standard value of $\sub{G}{isrf}$
could arise from shielding by low density gas from the surrounding molecular cloud.
The value of $\sub{G}{cr}$
is linked to the spectrum of cosmic rays and the
cosmic ray ionization rate \citep{web83,she04}. Variations in the initial
fraction of water ice can be caused by a different pre-collapse time: the
longer the pre-collapse time, the higher the fraction of water ice
(Table~\ref{t-03}).

Figure~\ref{f-04} shows the resulting abundance profiles (left panels)
and column densities (right panel) for water ice
and water vapour at a post-collapse time of $\sub{t}{post}=\unit[1]{Myr}$.
Because the focus of this paper is on the cold
chemistry, the column densities are not computed throughout the entire envelope,
but rather in the water freezeout-zone. This particularly affects the column density
of water vapour, which is abundant in the inner hot core \citep{boo03b}.

Variations of the ISRF shape the abundance profiles in the outer
envelope. The extinction threshold for the appearance of
the water vapour abundance plateau, $\sub{A}{V,f}$,
which is a result of the formation of a first monolayer of ice around the grains,
strongly depends on $\sub{G}{isrf}$. This behaviour, which has already been
described by \citet{hol09}, is a result of the attenuation of the ISRF, which
is accompanied by decreased photodesorption and photodissociation rates of
water ice and water vapour, respectively. At $\sub{A}{V}\gtrsim\sub{A}{V,f}$ the
desorption turns effectively from a first order (all water molecules on
the grain surface can desorb) into a zeroth order process (where only the
top layers contribute to desorption).

Variations in the CR-induced FUV field lead to changes in the abundance deep
in the envelope. Due to the reduction of the network to
$\gas\leftrightarrows\ice$, the water vapour abundance scales practically
1:1 with the $\sub{G}{cr}$. However, the abundance is too low to
considerably affect the water ice abundance.

Scaling the initial fraction of water ice has mostly an effect on the abundance
of water ice deep in the envelope. The temperatures of
$\sub{T}{dust}\gtrsim\unit[15]{K}$ inhibit the formation of water ice through
freeze-out of atomic oxygen. This has as a result that the abundance, and thus
the water ice column density, is already imprinted by the initial
conditions (cf.\ Sect.~\ref{s-repmodel}). The
effect of a higher binding energy of O is discussed in
Sect.~\ref{s-dis_ice}.

The column densities of water vapour are only mildly affected by parameter
variations within our parameter space. It should be noted that
a combination of low FUV fluxes from both the ISRF and the CR-induced
FUV field can indeed cause a reduction of this column density by a factor of $2-3$.
However, the ice column density remains hardly affected by the choice of FUV
photon fluxes, and envelope-averaged gas-to-ice ratios of $\sim10^{-4}$
within the water freeze-out zone are found. Variations of the initial fraction of
water ice $\sub{\xi}{\ice}$ only affect the water ice.
Since $\sub{\xi}{\ice}$ varies with the length
of the pre-collapse time, the observed water ice column densities
strongly depend on its initial abundance at the beginning of the post-collapse
phase.

\begin{table}[tb]
\centering
\caption{\label{t-04} Parameter space for the generation of
synthetic spectra.}
\begin{tabular}{l|cccl}
 \hline \hline Parameter & \multicolumn{3}{c}{Values} & Unit \\ \hline
 $\sub{t}{post}$ & $0.1$ & $1.0$ & & Myr \\
 $\sub{G}{isrf}$ & $10^{-2}$ & $10^{0}$& $10^{2}$ & \\
 $\sub{G}{cr}$ & $10^{-5}$& $10^{-4}$& $10^{-3}$ & \\
 $\sub{\vv}{r,env}$ & $0.0$& $\pm0.4$& $\pm0.8$ & km\,s$^{-1}$ \\
 $\beta$ & $0.4$& $0.8$& $1.2$ & km\,s$^{-1}$ \\
 $\sub{A}{V,j}$ & 0.0 & 5.0 & & mag \\
\hline
\end{tabular}
\tablefoot{Parameters are the post-collapse time $\sub{t}{post}$, the FUV fluxes from the
ISRF ($\sub{G}{isrf}$) and the CR-induced field ($\sub{G}{cr}$), the
maximum radial velocity ($\sub{\vv}{r,env}$; positive values represent
expansion), the turbulent broadening Doppler-$\beta$, and the
extinction threshold for a discontinuity in the Doppler-$\beta$ distribution
($\sub{A}{V,j}$).}
\end{table}

\begin{table*}[bt]
\centering
\caption{\label{t-05} Visual representation of the normalised Bayes Factor
$\sub{B}{p}(V_i)$ for each parameter $P$ (at its respective
grid point $V_i$).}
\begin{tabular}{l|cc|ccc|ccc|ccccc|ccc|cc}
\hline \hline Name & \multicolumn{2}{c|}{$\sub{t}{post}$ (Myr)} & \multicolumn{3}{c|}{$\logten(\sub{G}{isrf})$} & \multicolumn{3}{c|}{$\logten(\sub{G}{cr})$} & \multicolumn{5}{c|}{$\sub{\vv}{r,env}$ (km\,s$^{-1}$)} & \multicolumn{3}{c|}{$\beta$ (km\,s$^{-1}$)} & \multicolumn{2}{c}{$\sub{A}{V,j}$ (mag)} \\
& \tiny{$0.1$} & \tiny{$1.0$} & \tiny{$-2$} & \tiny{$0$} & \tiny{$+2$} & \tiny{$-5$} & \tiny{$-4$} & \tiny{$-3$} & \tiny{$-0.8$} & \tiny{$-0.4$} & \tiny{$0.0$} & \tiny{$+0.4$} & \tiny{$+0.8$} & \tiny{$0.4$} & \tiny{$0.8$} & \tiny{$1.2$} & \tiny{$0.0$} & \tiny{$5.0$} \\ \hline
L\,1527   & \EvA & \EvE & \EvE & \EvA & \EvA & \EvE & \EvA & \EvA & \EvE & \EvE & \EvE & \EvC & \EvA & \EvA & \EvD & \EvE & \EvE & \EvA \\
IRAS\,15398  & \nEv & \nEv & \EvE & \EvA & \EvA & \EvE & \EvA & \EvA & \EvE & \EvD & \EvA & \EvA & \EvA & \EvA & \EvE & \EvE & \EvE & \EvA \\ \hline
L1551-IRS5  & \EvA & \EvE & \EvE & \EvA & \EvA & \EvE & \EvA & \EvA & \EvA & \EvA & \EvA & \EvA & \EvE & \EvE & \EvB & \EvA & \EvE & \EvC \\
TMC1   & \EvA & \EvE & \EvE & \EvB & \EvC & \EvE & \EvE & \EvA & \nEv & \nEv & \nEv & \nEv & \nEv & \nEv & \nEv & \nEv & \EvE & \EvD \\
HH\,46   & \EvD & \EvE & \EvD & \EvD & \EvE & \EvE & \EvE & \EvA & \EvB & \EvC & \EvE & \EvE & \EvE & \EvE & \EvD & \EvC & \EvC & \EvE \\
TMC1A   & \EvC & \EvE & \EvE & \EvD & \EvA & \EvE & \EvE & \EvA & \EvD & \EvD & \EvE & \EvE & \EvE & \nEv & \nEv & \nEv & \EvE & \EvD \\
RCrA-IRS5  & \EvE & \EvB & \EvA & \EvA & \EvE & \EvE & \EvC & \EvA & \EvA & \EvA & \EvC & \EvE & \EvA & \EvA & \EvE & \EvA & \EvA & \EvE \\
TMR1   & \nEv & \nEv & \EvE & \EvE & \EvD & \EvE & \EvE & \EvC & \EvE & \EvE & \EvE & \EvE & \EvD & \nEv & \nEv & \nEv & \nEv & \nEv \\
L\,1489   & \EvD & \EvE & \EvE & \EvD & \EvA & \EvE & \EvE & \EvB & \EvD & \EvD & \EvE & \EvE & \EvE & \EvC & \EvD & \EvE & \EvE & \EvC \\
HH\,100-IRS  & \EvA & \EvE & \EvA & \EvE & \EvE & \EvA & \EvE & \EvA & \EvA & \EvA & \EvA & \EvC & \EvE & \EvA & \EvE & \EvA & \EvA & \EvE \\
RNO\,91   & \EvA & \EvE & \EvE & \EvA & \EvA & \EvE & \EvB & \EvA & \EvC & \EvD & \EvE & \EvE & \EvE & \EvD & \EvE & \EvE & \EvE & \EvC \\
\hline
\end{tabular}
\tablefoot{The bullets represent ranges of Bayes factors (Equation~\ref{e-13}),
namely\\
({\EvE}) $\logten[\sub{B}{p}(V_i)]\geq0.5$\\
({\EvD}) $-1.0\leq\logten[\sub{B}{p}(V_i)]<-0.5$\\
({\EvC}) $-1.5\leq\logten[\sub{B}{p}(V_i)]<-1.0$\\
({\EvB}) $-2\leq\logten[\sub{B}{p}(V_i)]<-1.5$\\
({\EvA}) $\logten[\sub{B}{p}(V_i)]<-2.0$\\
Generally, parameter values with $\logten[\sub{B}{p}(V_i)]\ll-1.0$ can
be rejected.}
\end{table*}

\subsubsection{Water Emission/Absorption Profiles}
\label{s-paramscan_gas}

In contrast to the observations of water ice, which are derived from the
attenuation of the light of a background source (and thus, only dependent on the
\textit{column} density along the line-of-sight), the spectrally resolved
water vapour lines allow us to model the underlying \textit{number} density
and velocity profile of the envelope. Due to the large beam
size of \textit{Herschel}, the observed spectra do not show the pencil beam
spectrum towards the core centre, but the major contribution actually
originates in lines-of-sight that intersect the envelope at impact parameters
of up to a few thousand au. Therefore, the spectrum has to be modelled through
radiative transfer tools to take into account the complex
interplay of excitation conditions, abundances, and velocities within the
sampled area on the sky. A simple absorption study will at best
only yield a lower limit on the water vapour column
of typically $\unit[10^{13}]{cm^{-2}}$, even for lines such as
the $p$-$\paraground$ line at $\unit[1113]{GHz}$ which are primarily
in absorption \citep{kri10}.

To demonstrate the influence of various parameters on the emission profiles,
we model the $o$-$\orthoground$ ground state line of
L1551-IRS5 and explore its dependence on FUV fluxes, post-collapse time, and velocity profile
with the aid of \textsc{Ratran} \citep{hog00}, following a similar approach as
\citet{mot13}. The initial abundances
are determined from our dark cloud model with a standard pre-collapse time
of $\sub{t}{pre}=\unit[0.1]{Myr}$. Figure~\ref{f-05} shows synthetic
spectra for four different parameter combinations.
For \textit{Model A}, the
post-collapse time is set to $\sub{t}{post}=\unit[1.0]{Myr}$,
and it assumes high FUV fluxes of $\sub{G}{isrf}=10^2$ and
$\sub{G}{cr}=10^{-3}$, a static envelope with zero
radial velocity, and a constant turbulent broadening, which is
characterised by Doppler-$\beta=\unit[0.4]{km\,s^{-1}}$.
The synthetic spectrum shows poor agreement with the observations.
In \textit{Model B}, the FUV fluxes are
reduced to $\sub{G}{isrf}=10^{0}$ and $\sub{G}{cr}=10^{-4}$. This results in
a shift of the abundance peak to lower extinctions.
In contrast to the line flux, which is in good agreement with the observations,
the skewed line profile cannot be reproduced.
Including expansion motions in \textit{Model C}
results in a skewed line and a good match of the absorption feature,
but a poor match to the peak emission. Only
a further decrease of the FUV photon fluxes result in good agreement with the
the observations (\textit{Model D}). This example highlights the
complex interplay of abundance structure and velocity profile to shape the
emission and absorption lines as observed with \textit{Herschel}.

\section{Comparison with Observations}
\label{s-comparison}

\subsection{Water Vapour}
\label{s-water_vapour}

\begin{figure*}[tb]
 \includegraphics{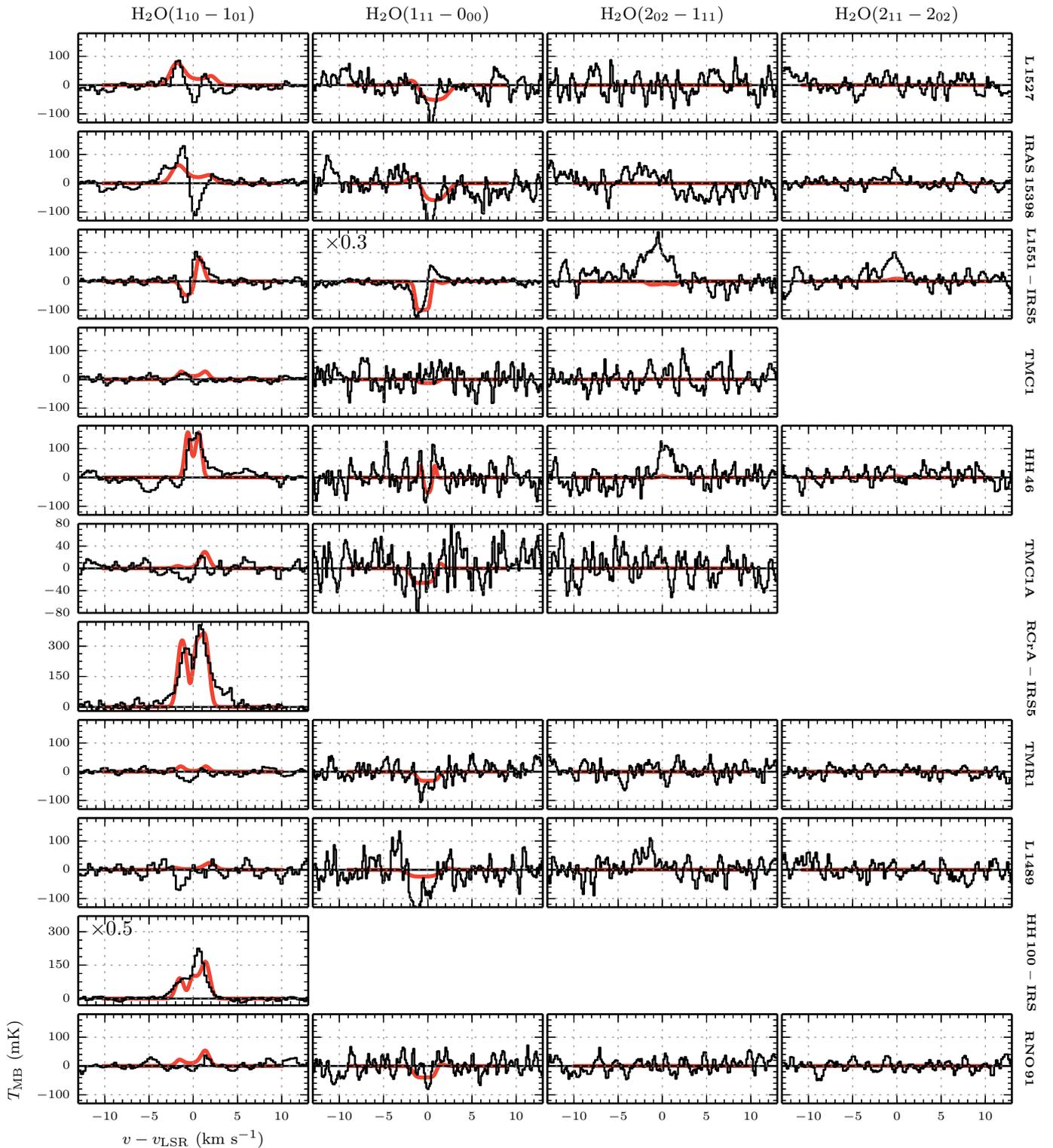}
 \caption{\label{f-06} Overview of the best fits of the
outflow- and continuum-subtracted \textit{Herschel} spectra
for protostellar cores in our sample. The spectra are shown in black, the
best-fits in red. We only fit the ortho and para ground state lines
(first and second column).}
\end{figure*}

The observed line profiles exhibit a remarkable diversity
(Fig.~\ref{f-01}), even though they all share
the same overall structure of the water vapour abundance profiles. As
shown in Sect.~\ref{s-paramscan_gas} (Fig.~\ref{f-05}),
the interplay of the abundance and velocity structure has a significant
impact on shaping the line profiles.

In the following parameter study, we vary parameters that actively influence
the shape of the water vapour abundance profile (the post-collapse time $\sub{t}{post}$,
and the FUV fluxes $\sub{G}{isrf}$ and $\sub{G}{cr}$). In addition,
the radial velocity distributions, namely the velocity centroid and the
Doppler broadening, are investigated. For the sake of simplicity, only
Hubble-like radial infall/expansion velocities are considered, which are
of the form
\begin{align}
 \sub{\vv}{r}(r) = \frac{r\,\sub{\vv}{r,env}}{\sub{r}{env}}
\end{align}
where $\sub{r}{env}$ is the envelope radius, and
$\sub{\vv}{r,env}$
the velocity at this point. Negative velocities mark infall, positive values
represent expansion. To account for possible discontinuities in the
Doppler-$\beta$ distribution, for which hints have been discovered
in both low- and high-mass cores \citep{her12,mot13}, various cases of
Doppler-$\beta$ distributions are tested. Firstly, a constant
Doppler-$\beta$ for the whole envelope is tested.
Secondly, an extinction threshold
$\sub{A}{V,j}$, which marks a jump in the Doppler-$\beta$
distribution, is introduced. Doppler-$\beta$ is only varied in regions
$\sub{A}{V}\geq\sub{A}{V,j}$, whereas in the outer regions
$\sub{A}{V}<\sub{A}{V,j}$ a constant
$\beta=\unit[0.2]{km\,s^{-1}}$ is chosen, which is motivated by the
presence of narrow absorption features in some sources (e.g., RCrA-IRS5).
Our parameter space is therefore defined by
$\theta=\left\{\sub{t}{post}, \sub{G}{isrf}, \sub{G}{cr}, \sub{\vv}{r,env}, \beta, \sub{A}{V,j}\right\}$.

The initial abundances are fixed after a pre-collapse time of
$\sub{t}{pre}=\unit[0.1]{Myr}$, since it has been shown in
Sect.~\ref{s-paramscan_gasice} that the choice of the pre-collapse time
(i.e., the inital abundance of water ice $\sub{\xi}{\ice}$)
does not considerably affect the water vapour abundance profiles in this study.

Owing to the line shape complexity and the plethora of parameters,
we do not attempt to find a singular ``best fit'' point in our parameter
grid. We rather aim to find general trends in the solutions, analysing the
influence of each parameter individually on the overall fit quality through
averaging over the values (i.e. marginalisation) of all other parameters.
This is pursued by adopting a Bayesian approach, which
is summarised in more detail in Appendix~\ref{s-bayes}. A parameter $p$ can
take $m$ values $V=\left\{V_1, \ldots, V_m\right\}$ (Table~\ref{t-04}). The overall
fit quality of our model with a particular value $V_i$ is quantified by
its \textit{evidence} $\sub{E}{p}(V_i)$. This number becomes only meaningful
when the evidences for all possible
parameter values $V$ are compared to each other. For each value $V_i$
we therefore normalize the evidence by defining the Bayes Factor as
$\sub{B}{p}(V_i)=\sub{E}{p}(V_i)/\mathrm{max}[\sub{E}{p}(V)]$. The parameter
value with maximised likelihood has therefore a
Bayes Factor of unity. Generally,
it is assumed that a parameter value can be rejected if
$\sub{B}{p}(V_i)\ll 0.1$. It should be noted that this approach is only
a relative comparison of the grid points in the parameter space, but does not
quantify the absolute quality of the fit.

The envelope models of \citet{kri12} do not take into account any deviation from
spherical symmetry (e.g., due to the presence of a disk) at scales
of $\lesssim\unit[300-500]{au}$. Therefore, we focus on the outer regions of
the envelope, and fit only ortho and para
ground state lines $\orthoground$ and $\paraground$. 

A summary of the parameters that yield the
best representation of the observed line profiles for all
sources is presented in Table~\ref{t-05}, and the spectra
are shown in Fig.~\ref{f-06}. In some cases the
overplotted best-fit spectra seem to be poor fits, but this
is owed due to the fact that we have chosen a coarse grid in our Bayesian analysis
plus an overall complexity in the source structure, which is heavily
simplified by our assumption of a radially symmetric envelope.
Instead of fitting every detail of the spectral lines, the goal of our
analysis is to find global trends, i.e., if a certain region
in the parameter space yields significantly better fits. In particular the
infall profiles of the Class~0 sources lack the deep absorption feature in the
$\orthoground$ line. \citet{mot13} showed that a detailed modelling of the
influence of the absorption against the outflow is able to recover the absorption
depth, but this kind of modelling is beyond the scope of this paper. Moreover,
L\,1551-IRS5 and HH\,46 show emission peaks in the higher excited lines, which
we attribute to the presence of structure in the inner regions
that is not part of our radially symmetric envelope model.

A few general trends for our sample of protostars can be derived. All sources
(except RCrA-IRS5) are characterised by a chemical age for water
of around $\unit[1]{Myr}$, but
in some, an age of $\unit[0.1]{Myr}$ cannot be ruled out.
Typically, our sources are exposed to low CR-induced FUV fields
($\sub{G}{cr}\lesssim 10^{-4}$). Mostly, we also find weak ISRF
($\sub{G}{isrf}\lesssim1$). In terms of velocity,
the two Class~0 sources are characterised by infall motions, and
Class~I sources show expansion motions of the envelope. However, due to
weak emission, the velocity field is hard to constrain in some
sources.
There seems to be no general trends for the turbulent broadening
(Doppler-$\beta$), but a few sources show better results when including
a discontinuity in the Doppler-$\beta$-distribution.

\subsection{Water Ice}
\label{s-res_ice}

As discussed in Sect.~\ref{s-paramscan_gasice}, the final abundance of water
ice in our models is practically independent of the FUV photon fluxes
over the parameter space considered, but strongly depends on the initial
oxygen abundances (namely, the fraction of water ice $\sub{\xi}{\ice}$, but also
the total abundance of oxygen $\sub{X}{O,SWaN}$ in the chemical network).
In our subsequent modelling, this parameter is taken from a dark cloud model
with different pre-collapse timescales $\sub{t}{pre}=0.01$, $0.1$ and
$\unit[1.0]{Myr}$ (Table~\ref{t-03}). Then, our
simplified chemistry is run for post-collapse time of $\sub{t}{post}=0.1$
and $\unit[1.0]{Myr}$. To determine the abundance from the column densities,
both the observed and simulated water ice column densities are divided
by the column density of hydrogen atoms in the water
freeze-out zone, $\sub{N}{H}^*$.

\begin{figure}[tb]
 \begin{center}
 \includegraphics{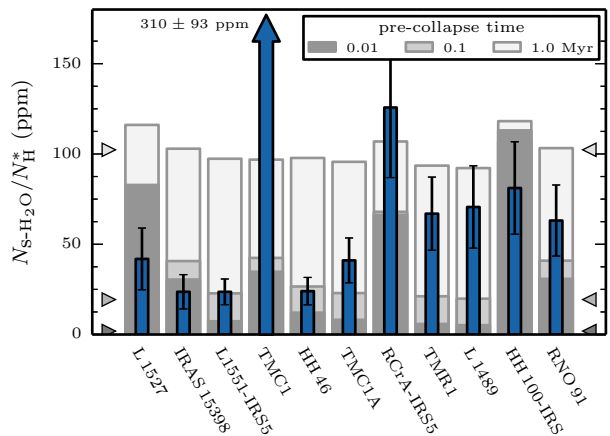}
 \caption{\label{f-07} Observed (blue) and modelled (grey)
 column-density-averaged water ice abundance ratios $\sub{N}{\ice}/\sub{N}{H}^*$ in
 the water freeze-out zone. The displayed models
 show the water ice abundances for pre-collapse times of 0.01,
 0.1, and $\unit[1.0]{Myr}$, followed by a post-collapse time of $\unit[1.0]{Myr}$.
 Another run with $\sub{t}{post}=\unit[0.1]{Myr}$ (not displayed here)
 reveals a marginal dependence
 of the abundances on the post-collapse time, with the results being the same
 as for $\sub{t}{post}=\unit[1.0]{Myr}$ within $\lesssim\unit[10-15]{ppm}$.}
 \end{center}
\end{figure}

We note that the abundances strongly vary with
pre-collapse time, but varying the post-collapse from 0.1 to
$\unit[1.0]{Myr}$ makes only a marginal difference. A comparison of
the observations and the modelled abundances
is found in Fig.~\ref{f-07}. In general, water ice abundances
range from $\sub{X}{\ice}=\unit[30-80]{ppm}$, except for TMC1 and
RCrA-IRS5. It should be noted that the observed
water ice column densities are always upper limits for the column density
that is part of the envelope. In some cases, a considerable amount of
water ice column density could originate from elsewhere along the line-of-sight
(e.g., foreground clouds, disks).
The exceptionally high water ice abundance in TMC1, e.g., might be
explained by the presence of a large disk \citep{har14}, which is intersected
by the line-of-sight towards the central protostar. In that case, the major fraction
of the absorbing water ice column in TMC1 would originate in the disk rather
than the envelope. The disks around TMC1A and TMR1, which are also
in the sample of \citet{har14}, seem to not affect the water ice abundances,
which lets us suggest that the line-of-sights towards these protostars are
pristine, and truly only through the envelope.

The modelled abundances show strong source-to-source variations in the
abundance of water ice for a short 
pre-collapse time, $\unit[0.01]{Myr}$. These variations become smaller with
increasing pre-collapse time,
and almost disappear for $\sub{t}{pre}=\unit[1.0]{Myr}$. To understand
this behaviour, one has to understand that in the prestellar core phase,
water ice slowly builds up through freeze-out of atomic oxygen. This
build-up is stopped in the post-collapse phase, where parts of the envelope
are too warm to allow freeze-out of atomic oxygen.
The observed
differences in the abundance structure on short pre-collapse timescales
therefore reflect the volume density and temperature structure of the
individual protostellar envelopes. In the envelopes of
L1551-IRS5, HH\,46, TMC1A, TMR1, and L\,1489, the bulk of material
in the post-collapse stage is found at $\sub{T}{dust}\gtrsim\unit[15]{K}$ where
water formation through freeze-out of oxygen is inhibited.
The ice column density in these sources is, therefore, almost completely
imprinted during the pre-collapse phase. Other sources
(L\,1527, IRAS\,15398, TMC1, RCrA-IRS5, RNO91, and HH\,100) have
a post-collapse envelope with a large atomic
oxygen freeze-out zone, and the transformation of oxygen to water ice can
continue throughout the post-collapse stage.

In contrast to older sources with abundances
$\gtrsim\unit[60]{ppm}$, young sources (with the exception of TMC1)
are generally characterised by abundances $\unit[20-50]{ppm}$. To speak of
an evolutionary effect would be an overinterpretation of the data,
but it is clear that the observed low abundances require
a rather short pre-collapse phase of $\lesssim\unit[0.1]{Myr}$.
This is an apparent contradiction to
the estimated lifetime of prestellar cores of $\sim\unit[0.5]{Myr}$
\citep{eno08}, which will be discussed in Sect.~\ref{s-dis_ice}.

\subsection{The Connection between Water Gas and Ice}

\begin{figure}[tb]
 \centering
 \includegraphics{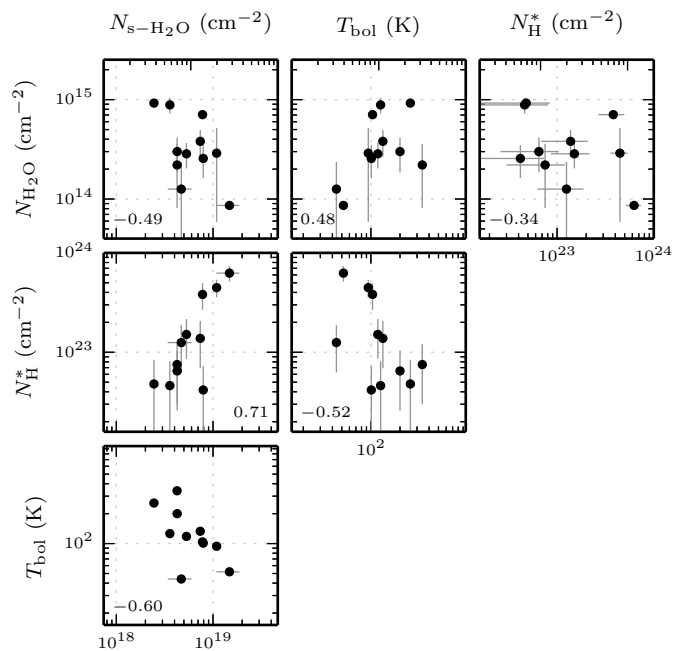}
 \caption{\label{f-08}Correlation of the bolometric temperature
 $\sub{T}{bol}$ and hydrogen column density $\sub{N}{H}$ with the column densities
 of water vapour $\sub{N}{\gas}$ and water ice $\sub{N}{\ice}$
 in the water freeze-out zone. The number in the corner
 depicts the Pearson sample correlation
 coefficient $r$.}
\end{figure}

\begin{table}
\caption{\label{t-06} Water vapour column densities and water gas-to-ice
ratios.}
\centering
\begin{tabular}{l|ee}
 \hline \hline Name & \multicolumn{1}{c}{$\sub{N}{\gas}$} & \multicolumn{1}{c}{$\sub{N}{\gas}/\sub{N}{\ice}$} \\
  & \multicolumn{1}{c}{($\unit[10^{14}]{cm^{-2}}$)} & \multicolumn{1}{c}{(10$^{-5}$)} \\ \hline
L\,1527         & 1.3,0.8 & 2.7,1.8 \\
IRAS\,15398     & 0.9,0.1 & 0.6,0.2 \\
L1551-IRS5      & 3.1,2.4 & 2.9,2.2 \\
TMC1            & 2.6,0.9 & 3.3,1.2 \\
HH\,46          & 7.3,0.6 & 9.3,1.2 \\
TMC1A           & 3.0,0.8 & 5.6,1.6 \\
RCrA-IRS5       & 8.2,0.7 & 23.0,2.7 \\
TMR1            & 3.5,0.4 & 4.7,0.5 \\
L\,1489         & 3.2,0.2 & 7.5,1.0 \\
HH\,100-IRS     & 9.3,2.3 & 38.0,10.3 \\
RNO\,91         & 2.2,1.4 & 5.2,3.3 \\
\hline
\end{tabular}
\end{table}

From the radiative transfer modelling in Sect.~\ref{s-water_vapour}
the likelihood-averaged column densities are determined.
Since the scatter in the derived values is mostly low,
the error bar is defined by the separation of the two column density
values in our discrete grid of parameters, which are closest to the best-fit
value. The water vapour column densities are on the order of
$\unit[10^{14}-10^{15}]{cm^{-2}}$.
These are significantly lower than the water ice column densities of
$\unit[10^{18}-10^{19}]{cm^{-2}}$, and the resulting gas-to-ice ratios are
on the order of $10^{-5}-10^{-4}$.
Correlations of water vapour and ice with
$\sub{N}{H}$ and $\sub{T}{bol}$ are depicted in Fig.~\ref{f-08}.
The only significant correlation is found for column densities of water ice
and hydrogen. Water vapour, on the other hand, is largely uncorrelated with
any of these parameters. This only reflects the nature of water ice as a
\textit{bulk} tracer, whereas water vapour rather traces the
\textit{surface layers} of the envelope. That obviously also
means that the gas-to-ice ratios, which vary by up to an order of magnitude from
source to source, are an intrinsic property of each individual source rather than
a global indicator.

\section{Discussion}
\label{s-discussion}

\subsection{Water Abundances with SWaN}
\label{s-dis_swan}

In Sect.~\ref{s-swan}, our simplified water network SWaN is introduced.
Through benchmarking with other chemical codes, the
chemical network can be limited to a small number of species and reaction
channels to understand the connection between water gas and ice,
and to reliably predict the
abundance profiles for both these species in the cold regions of the protostellar
envelope. The only species that needs to be added is atomic oxygen, which
takes over the role as a proxy for other oxygen bearing species in the
full water-chemistry network.

Simplifying the network to $\oxy\rightarrow\ice\rightarrow\gas\rightarrow\oxy$
in the outer envelope and to $\gas\leftrightarrows\ice$ further in, results
in equilibrium abundance profiles that are in good agreement with sophisticated
chemical networks. Photodesorption, photodissociation and freeze-out are
sufficient to explain the abundance structure of water vapour and water ice
in the water freeze-out zone.

\subsection{Water Ice}
\label{s-dis_ice}

A key ingredient in understanding the observed water ice column
densities is a low initial abundance of water ice after the
pre-collapse phase in concert with the presence of a freeze-out
barrier for atomic oxygen, caused by a relatively low binding
energy.

The initial abundances for the post-collapse phase are determined
through a dark-cloud model ($\sub{T}{dust}=\unit[10]{K}$,
$\sub{A}{V}=\unit[10]{mag}$,
$\sub{n}{H}=\unit[2\times10^4]{cm^{-3}}$; \citetalias{wal13}). The
formation of water ice is controlled by the freeze-out of atomic
oxygen. This freeze-out timescale, and thus the water ice formation
time scale, depends strongly on the product of the hydrogen number density, 
$\sub{n}{H}$, and the grain cross section per hydrogen atom,
$\sub{\sigma}{H}$ (Equation~\ref{e-10}), which in turn equals
$\sub{n}{gr} \sub{\sigma}{gr}$. Through a reduction of either $\sub{n}{gr}$
or $\sub{\sigma}{gr}$, or both, the formation of water ice can be
slowed down significantly. This could eradicate the apparent need
for a short pre-collapse time of $\sim\unit[0.1]{Myr}$ that was
found in our model, which conflicts with the observed prestellar
core lifetime of $\sim\unit[0.5]{Myr}$ \citep{eno08}. 

Our adopted grain abundance of $\sub{X}{gr}=6.5\times10^{-13}$
combined with a grain size of $\unit[0.1]{\mu m}$ corresponds to
$\sub{\sigma}{H} = \unit[2.0\times 10^{-22}]{cm^{2}}$
in our standard model. This
is almost an order of magnitude lower than the canonical value of
$\sub{\sigma}{H}=\unit[1.0\times10^{-21}]{cm^2}$
derived from diffuse cloud observations \citep[e.g.][]{pra83}, but
consistent with dense cores in which grains have grown to somewhat
larger sizes as predicted theoretically and found observationally
\citep[e.g.,][]{pag10,ste10}. Growth to even bigger sizes than
assumed here would be needed to bring the two timescales into
agreement.

The lifetime estimate of
\citet{eno08} refers to the dense prestellar cores with mean
densities above $\unit[2\times10^4]{cm^{-3}}$, to which their
millimetre continuum data were sensitive.
Water ice formation can already start to form in lower density,
translucent clouds with densities of a few $\unit[10^3]{cm^{-3}}$
\citep{whi01, cup07}. The time that the cloud spends in this low
density phase is unknown, but it would only add to the amount of
water ice after the pre-collapse phase, exacerbating the need for
a short dense, prestellar phase. Our short inferred pre-stellar
phase is in contrast with \citet{yil13b} who argued for a long
prestellar phase of at least $\unit[1]{Myr}$ at
$\sub{n}{H}=\unit[10^5]{cm^{-3}}$ to explain the absence of
gas-phase O$_2$ toward the NGC\,1333 IRAS4A protostellar
core. Since water ice is not observed directly toward this core,
it is not clear whether there is a similar discrepancy.

A key parameter is the freeze-out barrier for atomic oxygen,
which helps to maintain the low water ice abundance even after the
pre-collapse phase. Owing to the lack of laboratory measurements
for the binding energy of atomic oxygen on water ice or other
surfaces, a binding energy of $\sub{T}{b,O}=\unit[800]{K}$
\citep{tie82} was assumed. This results in a freeze-out
temperature of $\sub{T}{dust}\sim\unit[15]{K}$. Recent theoretical
work and laboratory data point to binding energies that could be
more of the order of $\sub{T}{b,O}=\unit[1500]{K}$ on amorphous
silicate or graphite surfaces \citep{ber08,he14}, for which the
freeze-out threshold temperature would be raised to around
$\sub{T}{dust}\sim\unit[35]{K}$. If such a high value would also
apply to the binding of O on water ice, about $2-5$ times more
material ($15-35\%$ of the column density within the water
freeze-out zone) would be found in a region where atomic oxygen
can still be converted into water ice, effectively increasing the
overall water abundance and thus requiring an even shorter
pre-stellar phase.

Efficient cosmic ray desorption in the chemical model of
\citet{cas02} helped to retain a considerable amount of atomic
oxygen in the gas phase. Their assumption of a low binding energy
($\sub{T}{b,O}=\unit[600]{K}$) decreased the cosmic ray desorption
timescale of atomic oxygen and resulted in desorption from the
grain before further processing. However, this is an effect that
we have not seen in our full chemical model benchmarking which
include cosmic ray desorption. Even for
$\sub{T}{b,O}\gtrsim\unit[800]{K}$, the residence time on the
surface is sufficient for hydrogenation reactions to convert
atomic oxygen into more tightly bound molecules such as OH or
H$_2$O.

As a result of this qualitative analysis, the apparent
contradiction between the modelled and observed prestellar core
lifetime remains. A possible way out of this is to assume that
after the prestellar core phase of $\sim\unit[0.5]{Myr}$ there are
mechanisms at play, which help to generate a water ice abundance
that \textit{resembles} the situation after a pre-collapse time of
$\sub{t}{pre}\sim\unit[0.1]{Myr}$. Our simple two-stage approach
with constant density and temperature profiles in each of the stages can
only be seen as a first approach, but should be followed by more
detailed modelling of a self-consistent protostellar core collapse, taking
into account the change in density, temperature and velocity
structure. A key player in influencing the abundance profiles of
water in a more realistic collapse scenario could be
episodic accretion. During bursts in accretion the source
luminosity can increase by $\sim2$ orders of magnitude
\citep{vor13}, which heats the envelope and can increase the
radius of the hot core by up to an order of magnitude
\citep{joh13}. Our simplified network is not designed to simulate
such a situation. Nevertheless, the temperature increase during
an accretion burst could trigger chemistry which drives oxygen
into other, more tightly bound species that are not included in
our simple chemistry network.

\subsection{Water Vapour Chemistry}
\label{s-dis_gas}

The water vapour emission lines
as observed with \textit{Herschel} turn out to be an
invaluable tracer for the FUV field and envelope kinematics \citep[cf.][]{mot13}.
Amongst our sample, 8/11 sources are characterised by low ISRF-FUV fluxes
$\sub{G}{isrf}\sim10^{-2}$. Assuming that the host
star-forming region of these sources is embedded in a standard ISRF
with $\sub{G}{isrf}=1$, a
water-free cloud with an extinction of $\sub{A}{V}\sim\unit[2-3]{mag}$
would be needed to provide the required attenuation. 
For example, all sources in the Taurus Molecular Cloud
(TMC1, TMC1A, L1551-IRS5, TMR1) show a trend towards $\sub{G}{isrf}=10^{-2}$.
This result is in agreement with the
findings of an extinction threshold for water ice in Taurus at
$\sub{A}{V}\sim\unit[3]{mag}$ \citep{smi93,tei99,whi88,whi01}.
Following Equation~(17) in \citet{hol09}, the
formation threshold for water ice can be stretched to
$\sub{A}{V}=\unit[2-3]{mag}$ for a tenuous ambient cloud with $\sub{G}{isrf}=1$
and $\sub{n}{H}\sim\unit[10^3]{cm^{-3}}$, which can remain almost devoid of
water due to the interplay of FUV field and low freeze-out rates. It would have an
extent of around $\unit[1]{pc}$, and extinction maps
of the Taurus Molecular Cloud suggest that this is not unreasonable
\citep{kai09}.

In some sources (HH46, RCrA-IRS5, HH100-IRS - and with less significance TMR1),
evidence for elevated ISRF FUV fields $\sub{G}{isrf}\gtrsim1$ is found, which
can originate in interaction of the outflow with the envelope or in an
photodissociation region (PDR). Particularly the latter seems to be the
case for HH100-IRS and RCrA-IRS5, which are known to be cocooned by the
extended strong radiation field of the PDR in the RCrA star-forming region.

When it comes to assessing the CR-induced secondary FUV field,
the observations are generally best represented
by $\sub{G}{cr}\lesssim10^{-4}$, with a trend towards $10^{-5}$. However,
recalling Equation~(\ref{e-02}), the photodesorption rate depends, amongst other
things, also on the product $\sub{\sigma}{gr}\,\sub{n}{gr}$. This product is
proportional to the inverse grain radius $a^{-1}$, i.e., grain growth will
result in a decreasing surface area per unit volume. Our measurements probe --
assuming that all other factors are well
known -- the product $\sub{\sigma}{gr}\,\sub{n}{gr}\,\sub{G}{cr}$ rather
than $\sub{G}{cr}$ alone. Therefore, low values of the CR-induced
FUV field in our models could indeed be the result of a reduced cosmic
ray flux deep inside dense cores \citep[e.g.,][]{pad13}. Alternatively, it could also be due to
dust growth to micron-sized particles deep in the envelope \citep[e.g.,][]{pag10,ste10}.
Unfortunately, these two effects cannot be distinguished by our analysis.

In terms of the post-collapse timescale, all sources show significant evidence
for $\sub{t}{post}=\unit[1.0]{Myr}$. In some sources,
$\sub{t}{post}=\unit[0.1]{Myr}$ can be ruled out, but this does not correlate with
evolutionary stage. In particular, the Class~0 source L\,1527 is best
represented by a post-collapse time of $\unit[1.0]{Myr}$, which is an order of
magnitude higher than the generally estimated lifetime of sources
at that evolutionary stage of $\sim\unit[0.1]{Myr}$ \citep{eva09}. But
that is only a contradiction at first glance. As seen earlier, the
initial molecular abundances that match the observations of \textit{water ice}
are best represented by a pre-collapse time of $\unit[0.1]{Myr}$
(Sect.~\ref{s-dis_ice}), and thus this was our choice for the determination
of the initial abundances. But we have argued that some mechanism, probably
episodic accretion, helps to reset the chemical clock in the inner regions
of the envelope to a situation that resembles a prestellar lifetime of
$\lesssim\unit[0.1]{Myr}$. But the outer envelope would not be affected by this,
and its chemical age would represent its true age. The water vapour profile in that
region is therefore governed by the cumulative timespan of the prestellar
stage ($\sim\unit[0.5]{Myr}$) plus the time the source has already spent
in Class~0 phase. Therefore, even the water vapour profiles of the
youngest Class~0 sources are characterised by
a chemical age of $\sim\unit[1.0]{Myr}$, rather than $\unit[0.1]{Myr}$.

\subsection{Kinematics}
\label{s-dis_kinematics}

Our analysis shows that water emission lines are not only a good
tracer for the FUV fluxes, but they also prove to be a sensitive
and invaluable tracer of the kinematics of protostellar envelopes. In the two
Class~0 sources we find strong evidence for infall motion. The best
fit spectrum seems to be a poor representation of the observed spectrum
(Fig.~\ref{f-06}). However, the focus of this paper is
on finding general trends (i.e., expansion vs.\ collapse) rather than determining
the true nature of the collapse. A free-fall collapse model, which
has been used by \citet{mot13} on sources with inverse P-Cygni profiles, does
indeed show better overall agreement with the observations. But quantifying
the nature of the infall is beyond the scope of this paper.

In our sample of Class~I sources, 7/9 sources are characterised by
expansion motion of the outer envelope. In 2/9 the kinematics remain mostly
unconstrained due to weak emission. This fits the picture of
envelope dispersal during this evolutionary stage \citep[e.g.,][]{kri12}.

Our inferred radial motions raise the question of how
reasonable the assumption of a static envelope is. Obviously, the determined
density, temperature, and velocity profiles are only snapshots of the
current situation, but have no information about their respective history.
At expansion velocities of the order of km/s as seen in L1551-IRS5 and
HH100-IRS, an envelope with an extent of $\sim\unit[10^4]{au}$ would have started
from a singular point less than $\unit[0.1]{Myr}$ ago. Therefore, this expansion
must have started fairly recent in time. However, our analysis shows that most
other Class~I sources show support for expansion, but it is impossible to
quantify its rate. Therefore, the assumption of a static envelope in the
post-collapse phase is our best approach, but it can only be seen as a
first step towards a self-consistent evolution model.

\subsection{The Oxygen Budget of Protostars}

\begin{figure}[tb]
 \includegraphics{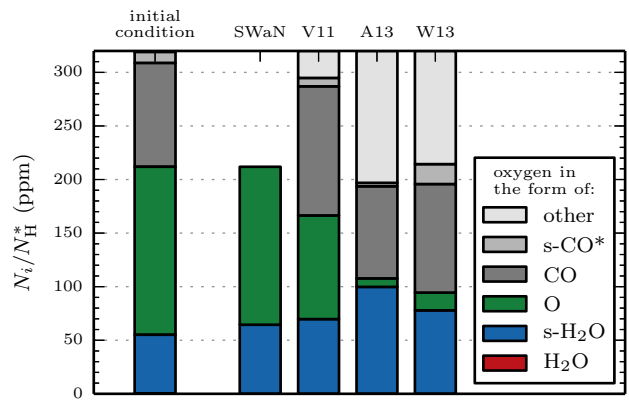}
 \caption{\label{f-09} Column-density-averaged
 abundance $N_i/\sub{N}{H}^*$ 
 of oxygen in water vapour ($\gas$), water ice
 ($\ice$), atomic oxygen ($\oxy$), carbon monoxide (CO),
 carbon-oxygen ices ($\mathrm{\textrm{s-}CO^*}$; these are s-CO, s-CH$_3$OH, s-H$_2$CO, which
 are the three most abundant molecules from the group of carbon-oxygen ices
 after the dark cloud phase), and other species in the
 water freeze-out zone for a post-collapse time of $\unit[0.1]{Myr}$.
 The initial conditions (Table~\ref{t-07})
 are shown in the left bar. This figure contains the results from
 the chemical models of \citetalias{vis11}, \citetalias{alb13}, and \citetalias{wal13},
 which are discussed in Appendix~\ref{s-benchmark}}
\end{figure}

The benchmarking of our simplified chemistry against full gas-grain
models also provides insight into the total oxygen budget of
protostellar envelopes. In Appendix~\ref{s-bench_obs} the predicted reservoirs
of oxygen in three other chemical networks are analysed. A snapshot of
the abundances after
$\sub{t}{post}=\unit[0.1]{Myr}$ (Fig.~\ref{f-09}) shows
that the two networks with
complex grain-surface chemistry (\citetalias{alb13}, \citetalias{wal13}) efficiently drain atomic
oxygen out of the system, and convert it into other species. This has
consequences for the oxygen budget puzzle and the nature
of the unidentified depleted oxygen \citep[UDO;][]{whi10}. 
Following the trail of the oxygen in these complex chemical networks
leads us to its possible hiding place. In \citetalias{wal13}, $\sim$30\% of the
oxygen is in species which are not listed in \citet{whi10}.
It predicts s-CO$_2$, s-H$_2$CO,
s-H$_2$O$_2$, O$_2$, and s-O$_2$ at abundances $\gtrsim 10^{-5}$,
and NO, s-NO, s-HNO and s-HCOOH and more complex
CHON species contributing $X\gtrsim10^{-6}$. Instead of having a single, large
oxygen reservoir, the UDO could actually be distributed amongst many
molecules -- both in icy and gaseous form.

\section{Conclusions and Summary}
\label{s-conclusion}

In this paper, \textit{Herschel}-HIFI observations of water vapour and
previously published observations of water ice column densities are
used to understand the connection between the water gas and ice in the
cold environment of protostellar envelopes.

\begin{enumerate}
    \item We develop a simple chemistry network, SWaN. The
water vapour and water ice abundances in the cold regions of pre- and
protostellar cores can be reliably
determined by only including freeze-out, photodesorption and photodissociation
for water vapour, water ice, and atomic oxygen (as a proxy for other
oxygen-bearing species). In the outer layers of protostellar envelopes
($\sub{A}{V}\lesssim\unit[3]{mag}$) the water abundance structure can be
determined by only considering the cycle
$\oxy\rightarrow\ice\rightarrow\gas\rightarrow\oxy$. At higher extinctions
(and also higher densities), where the freeze-out rather than photodissociation
dominates the destruction of water vapour, the network reduces to
$\gas\leftrightarrows\ice$.
    \item In cold, prestellar cores, the bulk of atomic oxygen is mostly
converted into
water ice on a timescale of $\unit[1-10]{Myr}$. In our sample of protostellar
cores we find that only $10-30\%$ of the oxygen is found in the form of
observed water ice. The key to model such a low abundance is a short prestellar core lifetime
of $\lesssim\unit[0.1]{Myr}$ together with the fact that the formation of
water ice is inhibited in protostellar envelopes in regions with
$\sub{T}{dust}\gtrsim\unit[15]{K}$ as a result of the assumed
oxygen binding energy of $\sub{T}{b,O}=\unit[800]{K}$. Use of a higher
binding energy reinforces our conclusion of a short pre-collapse
timescale. The
apparent contradiction of such short pre-collapse phases to observed lifetimes
of the dense phase of prestellar cores ($\sim\unit[0.5]{Myr}$)
can be circumvented by introducing a mechanism that efficiently reduces the
water ice abundance in the transition from a pre- to protostellar core.
One hypothesis is that of episodic accretion and its
effects on the abundance should be critically analysed.
    \item We find infall motions of the envelope for our two Class~0 sources,
and expansion motions in the majority of the Class~I sources. This is
consistent with infall during early stages, and envelope
dispersal during later stages. In addition, the water
vapour emission lines prove to be an excellent tracer for the FUV field strengths.
We find CR-induced FUV field strengths of $\sub{G}{cr}\lesssim10^{-4}$ for all
sources, and mostly low ISRF of $\sub{G}{isrf}\sim 10^{-2} - 10^0$ -- amongst
them three sources from the Taurus star-forming region. This attenuated ISRF
supports the finding of previous authors of an ice formation threshold at
$\sub{A}{V}\sim\unit[3]{mag}$.
    \item The finding of a low water ice abundance consistent with the observations
sheds light on the question of the question of the unidentified depleted oxygen (UDO).
Chemical modelling shows that upon introduction of an extended grains surface
chemistry network, oxygen can be distributed over many molecules (mostly ices).
Instead of a single, large oxygen reservoir, the UDO would consist of a
plethora of various species, which makes it hard for observers to track down
the complete oxygen budget in these sources.
\end{enumerate}

\begin{acknowledgements}
M.S.\ would like to thank Ted Bergin, Eric Keto and Paola Caselli for a
useful discussions and help with the
development of SWaN, and Coryn Bailer-Jones for assistance with the statistical
analysis. M.S\ acknowledges support from NOVA, the Netherlands Research School
for Astronomy. M.S.\ also acknowledges the use of astropy \citep{astropy},
NumPy, SciPy, and matplotlib \citep{matplotlib}. This research has made use of
NASA's Astrophysics Data System Bibliographic Services (ADS).

R.V. is supported by NASA through an award
issued by JPL/Caltech and by the National Science Foundation under grant 1008800.

C.W.\ acknowledges support from the European Union A-ERC grant 291141 CHEMPLAN
and financial support (via a Veni award) from the Netherlands Organisation
for Scientific Research (NWO).

Astrochemistry in Leiden is supported by the Netherlands Research
School for Astronomy (NOVA), by a Royal Netherlands Academy of Arts
and Sciences (KNAW) professor prize, by a Spinoza grant and grant
614.001.008 from the Netherlands Organisation for Scientific Research
(NWO), and by the European Community's Seventh Framework Programme
FP7/2007-2013 under grant agreement 238258 (LASSIE).

HIFI has been designed and built by a consortium of institutes and
university departments from across Europe, Canada and the United
States under the leadership of SRON Netherlands Institute for Space
Research, Groningen, The Netherlands and with major contributions
from Germany, France and the US. Consortium members are: Canada:
CSA, U.Waterloo; France: CESR, LAB, LERMA, IRAM; Germany: KOSMA,
MPIfR, MPS; Ireland, NUI Maynooth; Italy: ASI, IFSI-INAF,
Osservatorio Astrofisico di Arcetri-INAF; Netherlands: SRON, TUD;
Poland: CAMK, CBK; Spain: Observatorio Astron\'{o}mico Nacional
(IGN), Centro de Astrobiolog\'{i}a (CSIC-INTA). Sweden: Chalmers
University of Technology - MC2, RSS \& GARD; Onsala Space
Observatory; Swedish National Space Board, Stockholm University -
Stockholm Observatory; Switzerland: ETH Zurich, FHNW; USA: Caltech,
JPL, NHSC.
\end{acknowledgements}

\appendix

\section{Photodesorption and Spherically Averaged Extinction}
\label{a-av}

\begin{figure}[tb]
 \includegraphics{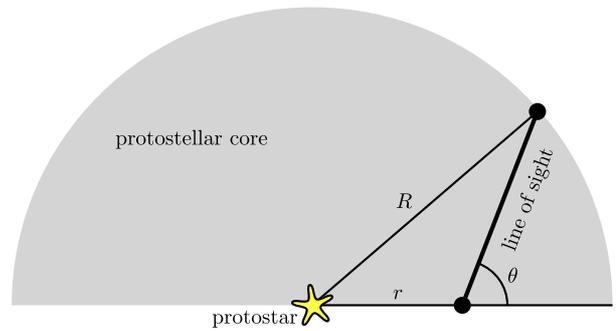}
 \caption{\label{f-10} The average extinction at
 arbitrary position into the cloud $r$ can be calculated by averaging over
 all lines of sight.}
\end{figure}

The photodesorption rate is closely connected to the rate at which FUV
photons impinge on the grain surface. An isotropic ISRF is characterised by
a flux of photons $\sub{F}{phot,0}$ through a unit surface
per unit time, which can be directly related to the isotropic
specific intensity $\sub{I}{phot,0}$ of FUV photons
\begin{align}
    \sub{F}{phot,0}&=2\pi\,\int_0^{\pi/2}\sub{I}{phot,0}\,\cos\theta\,\sin\theta\,\dd{\theta} = \pi \sub{I}{phot,0}.
\end{align}

Deeper into the envelope, the FUV field is attenuated according to
$\exp(-\gamma \sub{A}{V})$, where $\gamma$ is a pre-factor that depends on the
dominant FUV range for the process in question. At an arbitrary distance
$r$ from the core centre, the specific intensity of FUV photons is
isotropic, but rather depends on the azimuth angle
$\theta$ (Fig.~\ref{f-10}). The FUV photon density
can be calculated by
\begin{align}
    \sub{n}{phot}(r)&=\oint\,\frac{\sub{I}{phot,0}}{c}\,\ee{-\gamma\,\sub{A}{V}(r,\theta)}\,\dd{\Omega}\\[2ex]
    &=4\,\frac{\sub{F}{phot,0}}{c}\,\ee{-\gamma\,\Avbar(r)}.
\end{align}
In the last step, the directionally dependent extinction is replaced with the
spherically averaged extinction
\begin{align}
    \Avbar(r)&=-\frac{\ln\left[\int_0^\pi\,\frac{\sin\theta}{2}\,\ee{-\gamma\,\sub{A}{V}(r,\theta)}\dd{\theta}\right]}{\gamma}.
\end{align}

The rate at which FUV photons impinge on the surface of a single grain is given by
\begin{align}
    k &= 4\pi\,a^2\,\sub{F}{phot,0}\,\ee{-\gamma\,\sub{\bar{A}}{V}}.
\end{align}

Since photodesorption of ice and photodissociation of water vapour are
dominated by slightly different wavelength ranges, their respective
$\gamma$-factors are 1.8 and 1.7. This does produce slightly different
average extinctions. It should also be noted that $\Avbar$ differs
substantially from $\sub{A}{V}$, which denotes the extinction from the envelope edge
inwards in radial direction and is defined in Sect.~\ref{s-physmod}. For example, at the
envelope edge $\sub{A}{V}=\unit[0.0]{mag}$ we find a $\Avbar\sim\unit[0.4]{mag}$, at
$\sub{A}{V}=\unit[1.0]{mag}$ we get $\Avbar\sim\unit[1.9]{mag}$, and at
$\sub{A}{V}=\unit[10]{mag}$ we derive $\Avbar\sim\unit[11.8]{mag}$. These two
values converge towards the center.

\section{Chemical Network Benchmarking}
\label{s-benchmark}

\subsection{Other Chemical Networks}

\begin{figure}[tb]
 \includegraphics{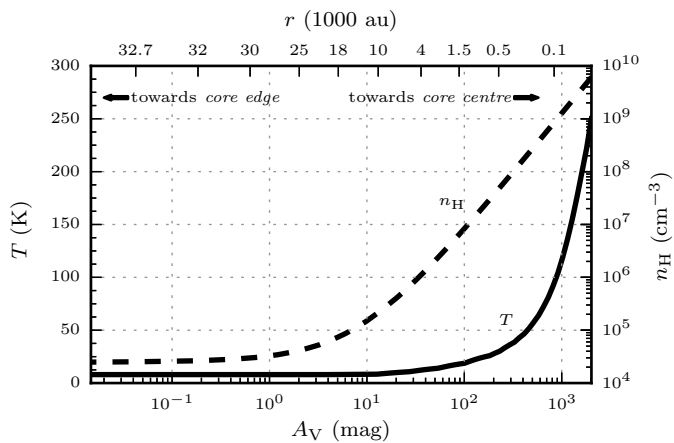}
 \caption{\label{f-11}The temperature and hydrogen density structure
 of NGC\,1333-IRAS4A as a function of radial extinction $\sub{A}{V}$.}
\end{figure}

\begin{figure}[tb]
 \includegraphics{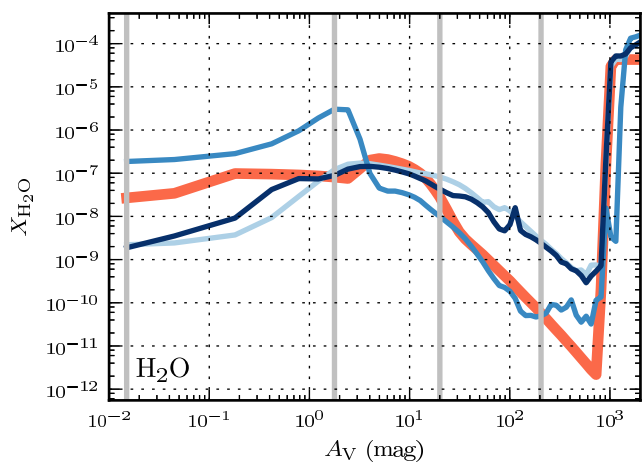}
 \caption{\label{f-12}Abundance structure of (from top to bottom)
 water vapour, water ice, atomic oxygen, and the sum of these three
 species at $t=\unit[0.1]{Myr}$.}
\end{figure}

Section~\ref{s-swan} introduced our simplified water network SWaN. To
assess its reliability, it is compared to three full chemical networks with
different databases, reaction channels, reaction mechanisms, and computational
details. The protostellar core NGC\,1333-IRAS4A is selected as a benchmark object,
since this source has been tested in detail against observations \citep{mot13}.

Section~\ref{s-swan} introduced our simplified water network SWaN. To
assess its reliability, it is compared to three full chemical networks with
different databases, reaction channels, reaction mechanisms, and computational
details. The protostellar core NGC\,1333-IRAS4A is selected as a benchmark object,
since this source has been tested in detail against observations \citep{mot13}.

\begin{itemize}
\item The network \citetalias{vis11} is based on \citet{vis11}, but fully updated
to the \textsc{Rate12} release of the UMIST Database for Astrochemistry
\citep[UDfA\footnote{http://www.udfa.net};][]{mce13}. Its grain-surface
chemistry is limited to the formation of H$_2$, H$_2$O, NH$_3$, CH$_4$, H$_2$S;
the latter four are formed simply through successive hydrogenation of the
respective heavy elements.

\item The network \citetalias{wal13} is based on \citet{wal13}
with additions from \citet{wal14}, which also employs the full
\textsc{Rate12} gas-phase chemistry. The network is supplemented with grain-surface
reactions from the Ohio State University network \citep[OSU;][]{gar08}, which
is much more extensive than the simple surface chemistry in \citetalias{vis11}. The grain-surface
network in \citetalias{wal13} forms both `simple' and `complex' ices, such as
H$_2$O, CH$_4$, CH$_3$OH, HCOOCH$_3$, and CH$_3$OCH$_3$. The ice chemistry
is treated as a single phase, i.e., the bulk and surface ice are not treated
separately. Grain-surface reactions occur via the Langmuir-Hinshelwood
mechanism only and reactive (or chemical) desorption is included with branching
ratio of 1\% \citep{gar07}.

\item The network \citetalias{alb13} is based on the deuterium chemistry model
of \citet{alb13}, which has been extended to include ortho-para chemistry
\citep{alb14a} and high-temperature reactions
\citep{alb14b}. It emerged from the protoplanetary
disk model of \citet{sem10}, who adapted the gas-grain model of \citet{gar08}.
\citetalias{alb13} has a similar grain-surface network as \citetalias{wal13}.
\end{itemize}

\begin{table}[tb]
 \caption{\label{t-07}Atomic/molecular abundances $X_i$ wrt.\ to
 hydrogen atoms at the start of the post-collapse phase.}
 \centering{
 \begin{tabular}{lc}
 \hline \hline atom/molecule & $X_i$ \\ \hline
 $\mathrm{H_{2}}$ & $\mathrm{5.00(-1)}$ \\
 $\mathrm{He}$ & $\mathrm{9.75(-2)}$ \\
 $\mathrm{H}$ & $\mathrm{1.98(-4)}$ \\
 $\mathbf{O}$ & $\mathbf{1.57(-4)}$ \\
 $\mathrm{CO}$ & $\mathrm{9.68(-5)}$ \\
 $\mathbf{\textbf{s-}H_{2}O}$ & $\mathbf{5.47(-5)}$ \\
 $\mathrm{N}$ & $\mathrm{4.54(-5)}$ \\
 $\mathrm{\textrm{s-}NH_{3}}$ & $\mathrm{1.24(-5)}$ \\
 $\mathrm{C}$ & $\mathrm{1.03(-5)}$ \\
 $\mathrm{\textrm{s-}CH_{4}}$ & $\mathrm{9.13(-6)}$ \\
 $\mathrm{N_{2}}$ & $\mathrm{7.69(-6)}$ \\
 $\mathrm{\textrm{s-}CO}$ & $\mathrm{6.02(-6)}$ \\
 $\mathrm{CH_{4}}$ & $\mathrm{3.75(-6)}$ \\
 $\mathrm{\textrm{s-}CH_{3}OH}$ & $\mathrm{2.45(-6)}$ \\
 $\mathrm{C_{3}}$ & $\mathrm{1.87(-6)}$ \\
 $\mathrm{\textrm{s-}H_{2}CO}$ & $\mathrm{1.84(-6)}$ \\
 $\mathrm{\textrm{s-}CH_{2}NH}$ & $\mathrm{1.03(-6)}$ \\
 $\mathbf{H_{2}O}$ & $\mathbf{4.26(-7)}$ \\
\ldots & \\ \hline
\end{tabular}
}
\tablefoot{Ice components are marked
 with the prefix ``s-''. The three species that are part of SWaN
 are marked in bold.}
\end{table}

All three networks contain several hundred species and several thousand
reactions. In addition to standard gas-phase chemistry, they all allow
freeze-out of neutral molecules onto cold dust grains. Desorption can occur
thermally, through direct cosmic ray heating, or through absorption of UV
photons. Also included in all three networks are photo-ionisation and
photodissociation, along with grain-surface formation of H$_2$.

All networks use the same set of parameters listed in Table~\ref{t-02},
together with a primary cosmic-ray ionisation rate of
$\sub{\zeta}{cr}=\unit[5.0\times10^{-17}]{s^{-1}}$. Desorption of ices is
treated as a zeroth-order process from only the top two monolayers
(Sect. \ref{s-abundances}). The abundances relative to hydrogen nuclei
$X_i\equiv n_i/\sub{n}{H}$ (Table~\ref{t-07}) are the same in all models,
and result from a dark cloud model ($T=\unit[10]{K}$, $\sub{n}{H}=\unit[2\times10^4]{cm^{-3}}$,
$\sub{A}{V}=\unit[10]{mag}$) at $\sub{t}{pre}=\unit[0.1]{Myr}$ \citep{wal13}
using the \textsc{Rate06} coefficients with the OSU grain surface network.
Table~\ref{t-07} highlights in bold the three species included in
SWaN: water vapour, water ice and atomic oxygen. Their cumulative
abundance, defined as $\sub{X}{O,SWaN}\equiv
X_\gas+X_{\ice}+X_\oxy$, is initially $\unit[212]{ppm}$\footnote{This value is
used for the benchmarking purposes. The difference to the
$\unit[270]{ppm}$ (Table~\ref{t-03}) that were used
in the main text of the paper originates in the use of different rate
coefficients (\textsc{Rate06} vs.\ \textsc{Rate12}).} out of the
available $\sub{X}{O,ISM}=\unit[320]{ppm}$ \citep{mey98}. Other
oxygen-bearing species like carbon monoxide or methanol are not
considered, which limits the amount of oxygen in SWaN to $\sim$66\% of
the abundance of volatile oxygen in the ISM.

The abundance profiles for water vapour, water ice, and atomic oxygen
(plus the sum of these three species) at $\sub{t}{post}=\unit[0.1]{Myr}$ are
presented in Fig.~\ref{f-12}. At all radii, the three full
networks agree to within a factor of 2 on the \textbf{water ice} abundance.
\citetalias{vis11} and \citetalias{wal13} show equally good agreement for
\textbf{water vapour}, but \citetalias{alb13} differs by up to
two orders of magnitude. This originates in the different treatment of
the desorption: In contrast to \citetalias{vis11} and \citetalias{wal13}, where
only molecules from the top layers can desorb, \citetalias{alb13} allows
desorption from the full ice mantle. SWaN recovers
the basic abundance structure for water ice and vapour to within the
uncertainties from the full networks, except for an underestimate of the ice
abundance at the very outer edge of the envelope.

The situation for \textbf{atomic oxygen} is more complicated, and the
predicted abundance structure of SWaN is less reliable. All networks agree on
a high O abundance at low $\sub{A}{V}$ due to photodesorption and
photodissociation of water and other oxygen-bearing species, but in the
shielded regions with higher densities, the oxygen abundance in SWaN strongly
deviates from the other networks. The reason for this
lies in the fact that the sophisticated chemical networks include
certain channels that allow atomic oxygen to react further to form other
species like O$_2$, CO, which are not included in SWaN. The bottom
panel of Fig.~\ref{f-12} shows the cumulative abundance
of all species present in SWaN. At high extinctions (and high densities),
the complex networks feature a net flow of oxygen into species
that are not part of SWaN. Hence, the abundance profile of atomic oxygen in SWaN
does not reflect the true profile in protostellar envelopes, but
rather acts as a oxygen reservoir that represents all other oxygen-bearing
species that are present in in the environment. \textit{Overall,
considering the limitations of the network, the agreement of both
the water vapour and water ice abundances with the complex chemical
networks is excellent.}

\subsection{Implications for Observations}
\label{s-bench_obs}

The modelled abundance profiles can be turned into observable column
densities by integrating over radius. Comparison to the total hydrogen
column density then gives the average line-of-sight column
density ratios, or `abundances', which can be compared to
observations. As discussed in Sect.~\ref{s-physmod}, our analysis is limited to the
water freeze-out zone, i.e., radii where
$\sub{T}{dust}\lesssim\unit[100]{K}$. Practically all water ice is found in
this region, which makes this point a convenient choice -- in
particular when comparing to observations of water ice.

\begin{table}[tb]
\caption{\label{t-08}Column density ratios of water vapour, water ice, and
 atomic oxygen in the water freeze-out region for the benchmark model.}
 \centering{
 \begin{tabular}{lccc}
 \hline \hline model & $N_\gas/\sub{N}{H}^*$ & $N_\ice/\sub{N}{H}^*$ & $N_\oxy/\sub{N}{H}^*$ \\
 & ($10^{-9}$) & ($10^{-5}$) & ($10^{-5}$) \\ \hline
  SWaN & 3.1 & 6.0 & 15.0 \\
  \citetalias{vis11} & 7.0 & 6.6 & 9.6 \\
  \citetalias{alb13} & 8.0 & 9.8 & 1.1 \\
  \citetalias{wal13} & 4.9 & 7.2 & 2.8 \\
 \hline
 \end{tabular}
}
\end{table}

In Table~\ref{t-08} the column density ratios are summarised for
all three species in the water freeze-out
zone for the benchmark NGC 1333 IRAS4A model. The agreement between
the chemical networks is good for water vapour (all are within a
factor of 2.5) and water ice (factor of 1.5). This
results in column-density-averaged gas-to-ice ratios of
$(3-11)\times10^{-5}$. On the other hand, atomic oxygen shows large
discrepancies of more than an order of magnitude. 

In Fig.~\ref{f-09} in the main text, the
column density ratios for all chemical networks alongside
the initial abundances from Table~\ref{t-07} are shown.
This figure illustrates how the oxygen budget
evolves in the protostellar phase. We distinguish between the chemical
species present in SWaN ($\gas$, $\ice$, $\oxy$), CO gas, carbon-oxygen-ices
($\mathrm{\textrm{s-}CO}$, $\mathrm{\textrm{s-}CH_3OH}$,
$\mathrm{\textrm{s-}H_2CO}$, which are the most abundant molecules from that
group after the pre-collapse phase), and other oxygen bearing species. Water
vapour is only present in trace amounts.

All chemical networks have only a small fraction of the oxygen in
water ice, about $\unit[50-100]{ppm}$, consistent with observations
\citep[e.g.][]{pon04,whi07}. One of the key findings of
Fig.~\ref{f-09} is the apparent drain of oxygen in the
sophisticated chemical networks towards species that are not part of
SWaN during the denser and warmer protostellar phase. Obviously, in
the simplified chemical network the abundance of oxygen has to be
conserved, since no other molecules are part of the network, and
atomic oxygen acts as the oxygen reservoir. However, in particular
the chemical networks \citetalias{alb13} and \citetalias{wal13} actively transform oxygen into other
species. The abundances of the CO gas and carbon-ices increase
marginally, but the atomic oxygen abundance drops from its initial
value of $\sim\unit[157]{ppm}$ to $\unit[5-15]{ppm}$. Other oxygen
bearing molecules , which in the beginning were only present in trace
amounts, make up $\sim\unit[106]{ppm}$ after only $\unit[10^5]{yr}$.

\section{Bayesian Analysis of the Water Emission Profiles}
\label{s-bayes}

The \textit{likelihood} that the observed data $D$
(a list of $N$ data points $\left\{y_k\right\}$) can be explained by model $M$ with
input parameters $\theta$ is, assuming a Gaussian error distribution, given by
\begin{align*}
    P(D | \theta, M) &= \left(\frac{1}{\sqrt{2\pi}\sigma}\right)^N\, \exp\left(-\frac{\chi^2}{2}\right)
\end{align*}
with
\begin{align*}
    \chi^2 = \sum_{k=1}^N \frac{(y_k-Y_k)^2}{\sigma^2}.
\end{align*}
$Y_k$ is a synthetic spectrum, which is the outcome of model $M$;
in our case a radiative transfer model with an abundance and velocity profile
determined through the parameters
$\theta=\left\{\sub{t}{post}, \sub{G}{isrf}, \sub{G}{cr}, \sub{\vv}{r,env}, \beta, \sub{A}{V,j}\right\}$.
The noise term $\sigma$ has contributions from the noise of the
observations $\sub{\sigma}{obs}$ (typically $\unit[10-30]{mK}$), and
an uncertainty in the synthetic RATRAN spectra of $\sim\unit[40]{mK}$. Different
weighting schemes (e.g., increasing uncertainties in the RATRAN spectra, but also
including velocity dependent terms) showed no influence on the general results.
Each grid point in the parameter space is assigned the minimum $\chi^2$ that is
reached when shifting the spectrum by $\pm\unit[0.4]{km\,s^{-1}}$ around
the individual systemic velocity, $\sub{\vv}{LSR}$ \citep{dis11,yil13a}
to account for uncertainties in the estimate of $\sub{\vv}{LSR}$.

In our analysis (Sect.~\ref{s-water_vapour}) we aim to find general
trends rather than a singular best fit point in our parameter grid $\varphi$ with
$\sub{N}{\varphi}$ grid points. To achieve this we
follow a Bayesian approach \citep{bai11}, where the overall fit quality
is assessed by the so-called \textit{evidence}, which is in the case of
a uniform prior distribution (i.e., all grid points are treated equally)
given by
\begin{align}
 \label{e-11}
    P(D|M)&=\frac{1}{\sub{N}{\varphi}} \sum_{\theta=\varphi} P(D|\theta, M).
\end{align}
The evidence is equivalent to the probability of the data $D$ given the model
$M$. The dependence on parameter $\theta$
has vanished due to marginalisation of the full parameter grid.

In our analysis, however, we are interested in variations of the fit quality
within the model upon variation of the value for particular parameter $P$.
Therefore, we modify our definition of the evidence in Equation~\ref{e-11}.
We define a new parameter space $\phi\equiv\varphi_{P=V_i}$
with $\sub{N}{\phi}$ grid points, where parameter $P$
takes value $V_i$. All other parameters remain free. We then calculate the
evidence by marginalising all free parameters, which is given by
\begin{align}
    \label{e-12}
    \sub{E}{p}(V_i) = \frac{1}{\sub{N}{\phi}}\,\sum_{\theta=\phi}P(D|\theta, M).
\end{align}
These evidences give an overall estimate of the fit quality, but their absolute
values are irrelevant. They only gain relevance when comparing to each other.
We therefore define the \textit{Bayes Factor} for each value $V_i$
of parameter $P$ as
\begin{align}
   \label{e-13}
    \sub{B}{p}(V_i) = \frac{\sub{E}{p}(V_i)}{\mathrm{max}[\sub{E}{p}(V)]}.
\end{align}
The parameter value with maximum likelihood has, therefore, a Bayes Factor of
unity. Other parameter values can also exhibit high likelihoods, but a value
is assumed to yield a poor representation if its Bayes Factor is $\ll0.1$.

This approach is illustrated by the following example: We analyse the
influence of the post-collapse timescale $\sub{t}{post}\equiv p$
with two values $V=\left\{\unit[0.1]{Myr}, \unit[1.0]{Myr}\right\}$. The
full parameter space is split up into two sub-spaces with
$\theta_i = \left\{\sub{t}{post}=V_i, \,\sub{G}{isrf} , \sub{G}{cr}, \sub{\vv}{r,env}, \beta, \sub{A}{V,j}\right\}$.
Marginalisation of all parameters except $\sub{t}{post}$ (Equation~\ref{e-12})
leads to evidences
$\sub{E}{p}(\unit[0.1]{Myr})$ and $\sub{E}{p}(\unit[1.0]{Myr})$. Let us now
assume, that maximum evidence is found for a post-collapse time of
$\unit[0.1]{Myr}$. We then get $\sub{B}{p}(\unit[0.1]{Myr})=1$ and
$\sub{B}{p}(\unit[1.0]{Myr})<1$, i.e., the most likely post-collapse time
is $\unit[0.1]{Myr}$. If we find
$\sub{B}{p}(\unit[1.0]{Myr})\ll0.1$, we can even assume that this is the
only reasonable solution, whereas we can reject $\sub{t}{post}=\unit[1.0]{Myr}$.

The summary for all parameters and all sources is shown in Table~\ref{t-05}.

\section{Observation}

The observating dates and IDs for each transition and source are listed
in Table~\ref{t-09}.

\begin{table}[htb]
\caption{\label{t-09}Source name, transition, observing date and observing
ID for all \textit{Herschel} spectra.}
\begin{tabular}{lccc}
\hline \hline Source Name & Transition & Obs.\ Date & Obs.\ ID \\ \hline
 L\,1527 & $1_{10}-1_{01}$ & 2010-03-21 & 1342192524 \\
 L\,1527 & $1_{11}-0_{00}$ & 2011-03-17 & 1342216335 \\
 L\,1527 & $2_{02}-1_{11}$ & 2010-08-18 & 1342203156 \\
 L\,1527 & $2_{11}-2_{02}$ & 2010-08-19 & 1342203214 \\
 IRAS\,15398 & $1_{10}-1_{01}$ & 2011-02-04 & 1342213732 \\
 IRAS\,15398 & $1_{11}-0_{00}$ & 2011-02-17 & 1342214414 \\
 IRAS\,15398 & $2_{02}-1_{11}$ & 2010-08-18 & 1342203165 \\
 IRAS\,15398 & $2_{11}-2_{02}$ & 2010-09-16 & 1342204795 \\
 L1551-IRS5 & $1_{10}-1_{01}$ & 2010-08-19 & 1342203194 \\
 L1551-IRS5 & $1_{11}-0_{00}$ & 2010-09-02 & 1342203940 \\
 L1551-IRS5 & $2_{02}-1_{11}$ & 2010-08-18 & 1342203153 \\
 L1551-IRS5 & $2_{11}-2_{02}$ & 2010-08-19 & 1342203219 \\
 TMC1 & $1_{10}-1_{01}$ & 2010-03-21 & 1342192526 \\
 TMC1 & $1_{11}-0_{00}$ & 2011-03-17 & 1342216336 \\
 TMC1 & $2_{02}-1_{11}$ & 2010-08-18 & 1342203155 \\
 HH\,46 & $1_{10}-1_{01}$ & 2010-05-10 & 1342196410 \\
 HH\,46 & $1_{11}-0_{00}$ & 2010-04-17 & 1342194785 \\
 HH\,46 & $2_{02}-1_{11}$ & 2010-04-18 & 1342195041 \\
 HH\,46 & $2_{11}-2_{02}$ & 2010-04-12 & 1342194560 \\
 TMC1A & $1_{10}-1_{01}$ & 2010-03-21 & 1342192527 \\
 TMC1A & $1_{11}-0_{00}$ & 2011-03-12 & 1342215969 \\
 TMC1A & $2_{02}-1_{11}$ & 2010-08-18 & 1342203154 \\
 RCrA-IRS5 & $1_{10}-1_{01}$ & 2011-03-10 & 1342215840 \\
 TMR1 & $1_{10}-1_{01}$ & 2010-03-21 & 1342192525 \\
 TMR1 & $1_{11}-0_{00}$ & 2010-09-02 & 1342203937 \\
 TMR1 & $2_{02}-1_{11}$ & 2010-08-18 & 1342203157 \\
 TMR1 & $2_{11}-2_{02}$ & 2010-08-19 & 1342203213 \\
 L\,1489 & $1_{10}-1_{01}$ & 2010-08-19 & 1342203197 \\
 L\,1489 & $1_{11}-0_{00}$ & 2010-09-02 & 1342203938 \\
 L\,1489 & $2_{02}-1_{11}$ & 2010-08-18 & 1342203158 \\
 L\,1489 & $2_{11}-2_{02}$ & 2010-08-19 & 1342203215 \\
 HH\,100-IRS & $1_{10}-1_{01}$ & 2011-03-10 & 1342215841 \\
 RNO\,91 & $1_{10}-1_{01}$ & 2010-09-29 & 1342205297 \\
 RNO\,91 & $1_{11}-0_{00}$ & 2011-02-17 & 1342214406 \\
 RNO\,91 & $2_{02}-1_{11}$ & 2010-09-14 & 1342204512 \\
 RNO\,91 & $2_{11}-2_{02}$ & 2010-09-16 & 1342204800 \\
\hline
\end{tabular}
\end{table}


\begin{thebibliography}{}
\bibitem[Aikawa et al.(2008)]{aik08} Aikawa, Y., Wakelam, V., Garrod, R.~T., \& Herbst, E.\ 2008, \apj, 674, 984
\bibitem[Aikawa et al.(2012)]{aik12} Aikawa, Y., et al.\ 2012, \aap, 538, A57
\bibitem[Albertsson et al.(2013)]{alb13} Albertsson, T., Semenov, D.~A., Vasyunin, A.~I., Henning, T., \& Herbst, E.\ 2013, \apjs, 207, 27
\bibitem[Albertsson et al.(2014a)]{alb14a} Albertsson, T., Indriolo, N., Kreckel, H., Semenov, D., Crabtree, K.~N., \& Henning, T.\ 2014, \apj, 787, 44
\bibitem[Albertsson et al.(2014b)]{alb14b} Albertsson, T., Semenov, D., \& Henning, T.\ 2014, \apj, 784, 39
\bibitem[Andersson \& van Dishoeck(2008)]{and08} Andersson, S., \& van Dishoeck, E.~F.\ 2008, \aap, 491, 907
\bibitem[Astropy Collaboration (2013)]{astropy} Astropy Collaboration 2013, \aap, 558, A33
\bibitem[Arasa et al.(2010)]{ara10} Arasa, C., Andersson, S., Cuppen, H.~M., van Dishoeck, E.~F., \& Kroes, G.-J.\ 2010, \jcp, 132, 184510
\bibitem[Bailer-Jones(2011)]{bai11} Bailer-Jones, C.~A.~L.\ 2011, \mnras, 416, 1163
\bibitem[Bergeron et al.(2008)]{ber08} Bergeron, H., Rougeau, N., Sidis, V., Sizun, M., Teillet-Billy, D., \& Aguillon, F.\ 2008, Journal of Physical Chemistry A, 112, 11921
\bibitem[Bergin et al.(1995)]{ber95} Bergin, E.~A., Langer, W.~D., \& Goldsmith, P.~F.\ 1995, \apj, 441, 222
\bibitem[Bergin et al.(2000)]{ber00} Bergin, E.~A., et al.\ 2000, \apjl, 539, L129
\bibitem[Bertin et al.(2012)]{ber12} Bertin, M., Fayolle, E.~C., Romanzin, C., et al.\ 2012, PCCP, 14, 9929
\bibitem[Bohlin et al.(1978)]{boh78} Bohlin, R.~C., Savage, B.~D., \& Drake, J.~F.\ 1978, \apj, 224, 132
\bibitem[Boogert et al.(2004)]{boo04} Boogert, A.~C.~A., et al.\ 2004, \apjs, 154, 359
\bibitem[Boogert et al.(2008)]{boo08} Boogert, A.~C.~A., et al.\ 2008, \apj, 678, 985
\bibitem[Boogert et al.(2011)]{boo11} Boogert, A.~C.~A., et al.\ 2011, \apj, 729, 92
\bibitem[Burke \& Brown(2010)]{bur10} Burke, D.~J., \& Brown, W.~A.\ 2010, PCCP, 12, 5947B
\bibitem[Boonman \& van Dishoeck(2003)]{boo03a} Boonman, A.~M.~S., \& van Dishoeck, E.~F.\ 2003, \aap, 403, 1003
\bibitem[Boonman et al.(2003)]{boo03b} Boonman, A.~M.~S., Doty, S.~D., van Dishoeck, E.~F., Bergin, E.~A., Melnick, G.~J., Wright, C.~M., \& Stark, R.\ 2003, \aap, 406, 937
\bibitem[van Broekhuizen et al.(2005)]{bro05} van Broekhuizen, F.~A., Pontoppidan, K.~M., Fraser, H.~J., \& van Dishoeck, E.~F.\ 2005, \aap, 441, 249
\bibitem[Bruderer et al.(2012)]{bru12} Bruderer, S., van Dishoeck, E.~F., Doty, S.~D., \& Herczeg, G.~J.\ 2012, \aap, 541, A91
\bibitem[Caselli et al.(2012)]{cas12} Caselli, P., et al.\ 2012, \apjl, 759, L37
\bibitem[Caselli et al.(2002)]{cas02} Caselli, P., Walmsley, C.~M., Zucconi, A., Tafalla, M., Dore, L., \& Myers, P.~C.\ 2002, \apj, 565, 344
\bibitem[Coutens et al.(2012)]{cou12} Coutens, A., et al.\ 2012, \aap, 539, A132
\bibitem[Coutens et al.(2013)]{cou13} Coutens, A., et al.\ 2013, \aap, 560, A39
\bibitem[Cuppen \& Herbst(2007)]{cup07} Cuppen, H.~M., \& Herbst, E.\ 2007, \apj, 668, 294
\bibitem[Chen et al.(1995)]{che95} Chen, H., Myers, P.~C., Ladd, E.~F., \& Wood, D.~O.~S.\ 1995, \apj, 445, 377
\bibitem[Cuppen et al.(2010)]{cup10} Cuppen, H.~M., Ioppolo, S., Romanzin, C., \& Linnartz, H.\ 2010, PCCP, 12, 12077
\bibitem[van Dishoeck \& Helmich(1996)]{dis96} van Dishoeck, E.~F., \& Helmich, F.~P.\ 1996, \aap, 315, L177
\bibitem[van Dishoeck et al.(2006)]{dis06} van Dishoeck, E.~F., Jonkheid, B., \& van Hemert, M.~C.\ 2006, Faraday Discussions, 133, 231
\bibitem[van Dishoeck et al.(2011)]{dis11} van Dishoeck, E.~F., et al.\ 2011, \pasp, 123, 138
\bibitem[van Dishoeck et al.(2013)]{dis13} van Dishoeck, E.~F., Herbst, E., \& Neufeld, D.~A.\ 2013, Chemical Reviews, 113, 9043
\bibitem[Dulieu et al.(2010)]{dul10} Dulieu, F., Amiaud, L., Congiu, E., Fillion, J.-H., Matar, E., Momeni, A., Pirronello, V., \& Lemaire, J.~L.\ 2010, \aap, 512, A30
\bibitem[Dulieu et al.(2013)]{dul13} Dulieu, F., Congiu, E., Noble, J., Baouche, S., Chaabouni, H., Moudens, A., Minissale, M., \& Cazaux, S.\ 2013, Scientific Reports, 3,
\bibitem[Enoch et al.(2008)]{eno08} Enoch, M.~L., Evans, N.~J., II, Sargent, A.~I., Glenn, J., Rosolowsky, E., \& Myers, P.\ 2008, \apj, 684, 1240
\bibitem[Evans et al.(2009)]{eva09} Evans, N.~J., II, et al.\ 2009, \apjs, 181, 321
\bibitem[Fraser et al.(2001)]{fra01} Fraser, H.~J., Collings, M.~P., McCoustra, M.~R.~S., \& Williams, D.~A.\ 2001, \mnras, 327, 1165
\bibitem[Garrod et al.(2007)]{gar07} Garrod, R.~T., Wakelam, V., \& Herbst, E.\ 2007, \aap, 467, 1103
\bibitem[Garrod et al.(2008)]{gar08} Garrod, R.~T., Weaver, S.~L.~W., \& Herbst, E.\ 2008, \apj, 682, 283
\bibitem[Gibb et al.(2004)]{gib04} Gibb, E.~L., Whittet, D.~C.~B., Boogert, A.~C.~A., \& Tielens, A.~G.~G.~M.\ 2004, \apjs, 151, 35
\bibitem[de Graauw et al.(2010)]{gra10} de Graauw, T., et al.\ 2010, \aap, 518, L6
\bibitem[Habing(1968)]{hab68} Habing, H.~J.\ 1968, \bain, 19, 421
\bibitem[Harsono et al.(2014)]{har14} Harsono, D., J{\o}rgensen, J.~K., van Dishoeck, E.~F., Hogerheijde, M.~R., Bruderer, S., Persson, M.~V., \& Mottram, J.~C.\ 2014, \aap, 562, A77
\bibitem[Hasegawa et al.(1992)]{has92} Hasegawa, T.~I., Herbst, E., \& Leung, C.~M.\ 1992, \apjs, 82, 167
\bibitem[He et al.(2014)]{he14} He, J., Jing, D., \& Vidali, G.\ 2014, PCCP, 16, 3493
\bibitem[Herpin et al.(2012)]{her12} Herpin, F., et al.\ 2012, \aap, 542, A76
\bibitem[Hiraoka et al.(1998)]{hir98} Hiraoka, K., Miyagoshi, T., Takayama, T., Yamamoto, K., \& Kihara, Y.\ 1998, \apj, 498, 710
\bibitem[Hogerheijde \& van der Tak(2000)]{hog00} Hogerheijde, M.~R., \& van der Tak, F.~F.~S.\ 2000, \aap, 362, 697
\bibitem[Hollenbach \& Tielens(1997)]{hol97} Hollenbach, D.~J., \& Tielens, A.~G.~G.~M.\ 1997, \araa, 35, 179
\bibitem[Hollenbach et al.(2009)]{hol09} Hollenbach, D., Kaufman, M.~J., Bergin, E.~A., \& Melnick, G.~J.\ 2009, \apj, 690, 1497H
\bibitem[Hunter (2007)]{matplotlib} Hunter, J.~D.\ 2007, Computing In Science \& Engineering, 9, 3, 90--95
\bibitem[Ioppolo et al.(2010)]{iop10} Ioppolo, S., Cuppen, H.~M., Romanzin, C., van Dishoeck, E.~F., \& Linnartz, H.\ 2010, PCCP, 12, 12065
\bibitem[Ioppolo et al.(2008)]{iop08} Ioppolo, S., Cuppen, H.~M., Romanzin, C., van Dishoeck, E.~F., \& Linnartz, H.\ 2008, \apj, 686, 1474
\bibitem[Johnstone et al.(2013)]{joh13} Johnstone, D., Hendricks, B., Herczeg, G.~J., \& Bruderer, S.\ 2013, \apj, 765, 133
\bibitem[J{\o}rgensen et al.(2002)]{jor02} J{\o}rgensen, J.~K., Sch{\"o}ier, F.~L., \& van Dishoeck, E.~F.\ 2002, \aap, 389, 908
\bibitem[Kainulainen et al.(2009)]{kai09} Kainulainen, J., Beuther, H., Henning, T., \& Plume, R.\ 2009, \aap, 508, L35
\bibitem[Keto et al.(2014)]{ket14} Keto, E., Rawlings, J., \& Caselli, P.\ 2014, arXiv:1403.0155 
\bibitem[Kristensen et al.(2010)]{kri10} Kristensen, L.~E., et al.\ 2010, \aap, 521, L30
\bibitem[Kristensen et al.(2012)]{kri12} Kristensen, L.~E., et al.\ 2012, \aap, 542, A8
\bibitem[McElroy et al.(2013)]{mce13} McElroy, D., Walsh, C., Markwick, A.~J., Cordiner, M.~A., Smith, K., \& Millar, T.~J.\ 2013, \aap, 550, A36
\bibitem[Meyer et al.(1998)]{mey98} Meyer, D.~M., Jura, M., \& Cardelli, J.~A.\ 1998, ApJ, 493, 222
\bibitem[Miyauchi et al.(2008)]{miy08} Miyauchi, N., Hidaka, H., Chigai, T., Nagaoka, A., Watanabe, N., \& Kouchi, A.\ 2008, Chemical Physics Letters, 456, 27
\bibitem[Mokrane et al.(2009)]{mok09} Mokrane, H., Chaabouni, H., Accolla, M., Congiu, E., Dulieu, F., Chehrouri, M., \& Lemaire, J.~L.\ 2009, \apjl, 705, L195
\bibitem[Mottram et al.(2013)]{mot13} Mottram, J.~C., van Dishoeck, E.~F., Schmalzl, M., Kristensen, L.~E., Visser, R., Hogerheijde, M.~R., \& Bruderer, S.\ 2013, \aap, 558, A126
\bibitem[Mottram et al.(2014)]{mot14} Mottram J.C., Kristensen L.E., van Dishoeck, E.F., et al, 2014, in press, to appear in A\&A
\bibitem[Oba et al.(2009)]{oba09} Oba, Y., Miyauchi, N., Hidaka, H., Chigai, T., Watanabe, N., \& Kouchi, A.\ 2009, \apj, 701, 464
\bibitem[{\"O}berg et al.(2009)]{obe09} {\"O}berg, K.~I.,Linnartz, H., Visser, R., \& van Dishoeck, E.~F.\ 2009, \apj, 693, 1209 
\bibitem[{\"O}berg et al.(2011)]{obe11} {\"O}berg, K.~I., Boogert, A.~C.~A., Pontoppidan, K.~M., van den Broek, S., van Dishoeck, E.~F., Bottinelli, S., Blake, G.~A., \& Evans, N.~J., II 2011, \apj, 740, 109
\bibitem[Pagani et al.(2010)]{pag10} Pagani, L., Steinacker, J., Bacmann, A., Stutz, A., \& Henning, T.\ 2010, Science, 329, 1622
\bibitem[Padovani et al.(2013)]{pad13} Padovani, M., Hennebelle, P., \& Galli, D.\ 2013, \aap, 560, A114
\bibitem[Pilbratt et al.(2010)]{pil10} Pilbratt, G.~L., et al.\ 2010, \aap, 518, L1
\bibitem[Prasad \& Tarafdar(1983)]{pra83} Prasad, S.~S., \& Tarafdar, S.~P.\ 1983, \apj, 267, 603
\bibitem[Przybilla et al.(2008)]{prz08} Przybilla, N., Nieva, M.-F., \& Butler, K.\ 2008, \apjl, 688, L103
\bibitem[Pontoppidan et al.(2004)]{pon04} Pontoppidan, K.~M., van Dishoeck, E.~F., \& Dartois, E.\ 2004, \aap, 426, 925 
\bibitem[Rieke \& Lebofsky(1985)]{rie85} Rieke, G.~H., \& Lebofsky, M.~J.\ 1985, \apj, 288, 618
\bibitem[Semenov et al.(2010)]{sem10} Semenov, D., et al.\ 2010, \aap, 522, A42
\bibitem[Shen et al.(2004)]{she04} Shen, C.~J., Greenberg, J.~M., Schutte, W.~A., \& van Dishoeck, E.~F.\ 2004, \aap, 415, 203
\bibitem[Snell et al.(2000)]{sne00} Snell, R.~L., et al.\ 2000, \apjl, 539, L101
\bibitem[Smith et al.(1993)]{smi93} Smith, R.~G., Sellgren, K., \& Brooke, T.~Y.\ 1993, \mnras, 263, 749
\bibitem[Steinacker et al.(2010)]{ste10} Steinacker, J., Pagani, L., Bacmann, A., \& Guieu, S.\ 2010, \aap, 511, A9
\bibitem[Teixeira \& Emerson(1999)]{tei99} Teixeira, T.~C., \& Emerson, J.~P.\ 1999, \aap, 351, 292
\bibitem[Tielens \& Hagen(1982)]{tie82} Tielens, A.~G.~G.~M., \& Hagen, W.\ 1982, \aap, 114, 245
\bibitem[Tielens \& Allamandola(1987)]{tie87} Tielens, A.~G.~G.~M., \& Allamandola, L.~J.\ 1987, Interstellar Processes, 134, 397
\bibitem[Visser et al.(2011)]{vis11} Visser, R., Doty, S.~D., \& van Dishoeck, E.~F.\ 2011, \aap, 534, A132
\bibitem[Walsh et al.(2013)]{wal13} Walsh, C., Millar, T.~J., \& Nomura, H.\ 2013, \apjl, 766, L23
\bibitem[Walsh et al.(2014)]{wal14} Walsh, C., Millar, T.~J., Nomura, H., Herbst, E., Widicus Weaver, S., Aikawa, Y., Laas, J.~C., \& Vasyunin, A.~I.\ 2014, \aap, 563, A33
\bibitem[Webber \& Yushak(1983)]{web83} Webber, W.~R., \& Yushak, S.~M.\ 1983, \apj, 275, 391
\bibitem[Vorobyov et al.(2013)]{vor13} Vorobyov, E.~I.,DeSouza, A.~L., \& Basu, S.\ 2013, \apj, 768, 131
\bibitem[Whittet et al.(1988)]{whi88} Whittet, D.~C.~B., Bode, M.~F., Longmore, A.~J., Adamson, A.~J., McFadzean, A.~D., Aitken, D.~K., \& Roche, P.~F.\ 1988, \mnras, 233, 321
\bibitem[Whittet et al.(2001)]{whi01} Whittet, D.~C.~B., Gerakines, P.~A., Hough, J.~H., \& Shenoy, S.~S.\ 2001, \apj, 547, 872
\bibitem[Whittet et al.(2007)]{whi07} Whittet, D.~C.~B., Shenoy, S.~S., Bergin, E.~A., Chiar, J.~E., Gerakines, P.~A., Gibb, E.~L., Melnick, G.~J., \& Neufeld, D.~A.\ 2007, \apj, 655, 332
\bibitem[Whittet(2010)]{whi10} Whittet, D.~C.~B.\ 2010, \apj, 710, 1009
\bibitem[Whittet et al.(2013)]{whi13} Whittet, D.~C.~B., Poteet, C.~A., Chiar, J.~E., Pagani, L., Bajaj, V.~M., Horne, D., Shenoy, S.~S., \& Adamson, A.~J.\ 2013, \apj, 774, 102
\bibitem[Y{\i}ld{\i}z et al.(2013a)]{yil13a} Y{\i}ld{\i}z, U.~A., et al.\ 2013a, \aap, 556, A89
\bibitem[Y{\i}ld{\i}z et al.(2013b)]{yil13b} Y{\i}ld{\i}z, U.~A., et al.\ 2013b, \aap, 558, A58
\bibitem[Zasowski et al.(2009)]{zas09} Zasowski, G., Kemper, F., Watson, D.~M., Furlan, E., Bohac, C.~J., Hull, C., \& Green, J.~D.\ 2009, \apj, 694, 459
\end{thebibliography}
\end{document}